\documentclass[twocolumn,reprint,amsmath,amssymb,aps,prl]{revtex4-1} \usepackage[colorinlistoftodos]{todonotes}
\usepackage{amsmath}
\usepackage{amssymb}
\usepackage{float}
\usepackage{graphicx}
\usepackage{color}
\usepackage{bm}
\usepackage{soul}
\usepackage{subfig}
\usepackage[export]{adjustbox}
\usepackage{caption}
\begin{document}
\newcommand{\beq}{\begin{equation}}
\newcommand{\eeq}{\end{equation}}
\newcommand{\bea}{\begin{eqnarray}}
\newcommand{\eea}{\end{eqnarray}}
\newcommand{\be}{\begin{equation}}
\newcommand{\ee}{\end{equation}}
\newcommand{\ba}{\begin{eqnarray}}
\newcommand{\ea}{\end{eqnarray}}
\newcommand{\nn}{\nonumber}
\def\mb{\mathbf}
\newcommand{\de}[1]{\textcolor{blue}{#1}}
\newcommand{\pa}[1]{\textcolor{blue}{#1}}
\captionsetup{justification=raggedright,singlelinecheck=false}

\title{
$T$-matrix approach to the phonon-mediated Casimir interaction
 }

\author{Andrei I. Pavlov}
\affiliation{Institute for Theoretical Solid State Physics, Leibniz-Institut f\"ur Festk\"orper-
und Werkstoffforschung IFW-Dresden, D-01169 Dresden, Helmholtzstraße 20,
Germany}
\author{Jeroen van den Brink}
\affiliation{Institute for Theoretical Solid State Physics, Leibniz-Institut f\"ur Festk\"orper-
und Werkstoffforschung IFW-Dresden, D-01169 Dresden, Helmholtzstraße 20,
Germany}
\author{Dmitri V. Efremov}
\affiliation{Institute for Theoretical Solid State Physics, Leibniz-Institut f\"ur Festk\"orper-
und Werkstoffforschung IFW-Dresden, D-01169 Dresden, Helmholtzstraße 20,
Germany}
\begin{abstract}

We develop a theory of the phonon mediated Casimir interaction between two point-like impurities, which is based on the single impurity scattering $T$-matrix approach.
We show that the Casimir interaction at $T=0$ falls off as a power law with the distance between the impurities.
We find that the power in the weak and unitary phonon-impurity scattering limits differs, and we relate the power law to the low-energy properties of the single impurity scattering $T$-matrix.
In addition, we consider the Casimir interaction at finite temperature and show that at finite temperatures the Casimir interaction becomes exponential at large distances.

\end{abstract}

\maketitle

\section{Introduction}
The Casimir interaction is the fundamental physical phenomenon, which emerges due to modifications of vacuum 
fluctuations by boundaries in a confined area of space \cite{Casimir1948, Dzyaloshinskii, Munday}. It arises in many fields of physics \cite{Mostepanenko, rev1, French2010}, including condensed matter  \cite{Volovik, cond4, cond2}. Recently, substantial interest in the Casimir effect has revived
as a result of the significant progress in the experimental techniques in cold atoms \cite{Moritz2003, Moritz2005, Tolra2004, Kinoshita2004, Catani2012, O'Hara2000, Kohl2005}. Several setups were proposed for studying the Casimir interaction in cold atoms. \cite{Volosniev, Wachter2007, Petk}. Among them is the Luttinger liquid in one-dimensional (1D) atomic gases \cite{Xianlong2002, Recati2003, Gleisberg2004, Tokatly2004, Astrakharchik2004}, in which phonons may mediate the Casimir interaction between two impurity atoms \cite{Recati2005, Schecter2014}. 
The Casimir interaction for this realization was studied in two limits in one dimension: by Recati \textit{et al.} in the static \cite{Recati2005} and by Schecter and Kamenev in the dynamic limit  \cite{Schecter2014}. 
It was shown, that the Casimir interaction falls off with the distance $r$ between the impurities as $r^{-1}$ in the static limit of impurities in the Luttinger liquid of fermions with repulsion, and as $r^{-3}$ in the dynamic one. Later on, the authors of \cite{Pavlov2018} examined how the Casimir interaction evolves with the increase of the mass of the impurities and showed that the scaling of the Casimir interaction continuously changes from $r^{-3}$ to $r^{-1}$ for dynamic impurities if the mass of the impurities becomes infinitely large. 
The observability of the Casimir effect in 1D cold atoms was checked in \cite{Recati2005,Schecter2014}. 
It was predicted there that the magnitude of the Casimir interaction in cold atoms is within the experimentally accessible range.

In the present paper, we extend the theory of the phonon mediated Casimir interaction to two and three dimensional systems at finite temperatures. We consider a model of two impurity atoms having different masses (e.g. isotopes), embedded in a lattice.  The scattering of the lattice phonons on the impurities gives rise to Casimir forces between the impurities.
We investigate the evolution of the Casimir interaction in the full range of the scattering amplitude starting from the weak phonon-impurity scattering to the unitary limit.  
In order to regularize the intrinsic infra-red and ultraviolet divergences, the impurity scattering $T$-matrix approach is used. Such an approach was used in earlier works in different contexts \cite{Engman1972,Klein1969, Halperin1975, Kayanuma1985, Rueff1977, Polishchuk1997,Scardicchio2005, Emig2007, Kenneth2008, Shytov2009, Jiang2019}.  

The rest of the paper is organized as follows. We describe the model for dynamical impurities. Then we derive the Casimir interaction in terms of the single-impurity $T$-matrix for the considered system. Using the general properties of the $T$-matrix, we consider the Casimir interaction in dimensions $D=1-3$. Then, we evaluate the effect of temperature. Finally, we consider a model for static impurities in an external potential. We conclude with a discussion of the obtained results.

\section{Dynamical impurities}
We consider acoustic phonons 
 which are described by a standard Hamiltonian (e.g. \cite{Atland}): 
\begin{equation}
\label{eq.H0} \hat{H}_0 = \sum_{\mb{k}} \big[\pi(\mb{k})\bar{\pi}(\mb{k})+\omega^2_{\mb{k}}\varphi(\mb{k})\bar{\varphi}(\mb{k})\big],
\end{equation}
where the bosonic fields are:
\beq \nonumber \pi(\mb{x})=\frac{\textit{i}}{\sqrt{V}}\sum_{\mb{k}} \sqrt{\frac{\omega_{\mb{k}}}{2}}\left[ b_{\mb{k}}e^{\textit{i}\mb{kx}}-b^{\dagger}_{\mb{k}}e^{-\textit{i}\mb{kx}} \right], \eeq
\beq \nonumber \varphi(\mb{x})=\frac{1}{\sqrt{V}}\sum_{\mb{k}} \sqrt{\frac{1}{2\omega_{\mb{k}}}}\left[ b_{\mb{k}}e^{\textit{i}\mb{kx}}+b^{\dagger}_{\mb{k}}e^{-\textit{i}\mb{kx}} \right], \eeq
with linear dispersion $\omega_{\mathbf{k}} = c|\mathbf{k}|$. Here $b_{\mb{k}}, b^{\dagger}_{\mb{k}}$ are phonon annihilation and creation operators, $V$ is the volume of the system, and $c$ is the sound velocity. We put $\hbar=1$ in the paper.

The simplest form of the interaction of the phonons with an impurity is the bilinear form of the field operators, i.e. $\pi \bar{\pi}$ and $\varphi \bar{\varphi}$. 
The term $\pi \bar{\pi}$ describes a perturbation in the kinetic energy and gives a dominant contribution to the Casimir interaction in systems with mobile impurities which are moving coherently with the media \cite{Schecter2014}.  Hereafter we refer to these type of impurities as dynamic impurities. The term $\varphi \bar{\varphi}$ characterizes a perturbation in the potential energy of immobile impurities. Therefore we label them as static impurities. 
The origin of these terms will be discussed below with the example of a model of two impurity atoms embedded in a lattice  (see Appendix A). In addition, the model can also describe the interaction between two impurities in the Luttinger liquid in one dimension as shown in \cite{Schecter2014}. 

In this section, we consider the $\pi \bar{\pi}$ impurity-phonon interaction for two impurities. We consider the case when the impurities are much slower than phonons. In this limit, the interaction of phonons with two impurities located at the given time at the coordinates $-\mb{r}/2$ and $\mb{r}/2$ can be written as   
\bea \nn  \hat{H}_{int}=-g\Bigg( \left.\pi\big (\mb{x})\bar{\pi}\big(\mb{x}\big)\right\vert_{\mb{x}=-\frac{\mb{r}}{2}}+\left.\pi\big (\mb{x})\bar{\pi}\big(\mb{x}\big)\right\vert_{\mb{x}=\frac{\mb{r}}{2}}\Bigg). \\
\label{v1v2}
\eea
Here $g$ is the interaction constant. The requirement of the positiveness of the kinetic energy leads to $g\leq 1$. It was shown in \cite{Pavlov2018} that $g=1$ corresponds to the limit of infinite mass of impurities in the lattice model (see Appendix A).

We define the Green's functions on the Matsubara axis ($\omega_n=2\pi Tn$) at temperature $T$ for non-interacting bosons as \cite{Abrikosov1975}:
\beq \label{eq.Green.function.definition} G^{(0)}\!\!(\mb{x},\mb{x}',\omega_n)\!=\!-\int_{0}^{\,\frac{1}{T}}\!\!\!\!d\tau e^{-\textit{i}\omega_n \tau}\langle T_{\tau}\big(\pi(\mb{x},\tau)\bar{\pi}(\mb{x}'\!,0)\big) \rangle. \eeq

For the calculations of the Casimir interaction, it is enough to find the Green's functions taken at the coordinates of the impurities $\pm\mathbf{r}/2$. For the sake of simplicity, we use the following notation:
\bea \nonumber G^{(0)}_{r}(\omega_n)&\equiv &G^{(0)}\big(+\frac{\mb{r}}{2},-\frac{\mb{r}}{2},\omega_n\big)=G^{(0)}\big(-\frac{\mb{r}}{2},+\frac{\mb{r}}{2},\omega_n\big), \\
\nn G^{(0)}(\omega_n)&\equiv &G^{(0)}\big(+\frac{\mb{r}}{2},+\frac{\mb{r}}{2},\omega_n\big)=G^{(0)}\big(-\frac{\mb{r}}{2},-\frac{\mb{r}}{2},\omega_n\big).\\
\label{Green_r_0}
 \eea
The explicit expressions for these Green's functions read \citep{Pavlov2018}
\beq
G^{(0)}_{\mb{r}}(\omega_n)=\int\frac{d^D\mb{k}}{(2\pi)^D}\left[1-\frac{\omega^2_{n}}{\omega_n^2+\omega^2_{\mb{k}}}\right]e^{-\textit{i}\mb{kr}}
\label{eq.GreenR}
\eeq
and
\beq
G^{(0)}(\omega_n)=\int\frac{d^D\mb{k}}{(2\pi)^D}\left[1-\frac{\omega^2_{n}}{\omega_n^2+\omega^2_{\mb{k}}}\right],
\label{eq.Green0}
\eeq
respectively. Note that $G^{(0)}(\omega_n)$ is formally divergent in the ultra-violet limit. It can be regularized considering a lattice model.

\section{The Casimir interaction} 
The starting point is the derivation of the thermodynamic potential of the system of phonons interacting with two impurities located at points $\pm \mathbf{r}/2$. We employ the well-known relation between the derivative of the thermodynamic potential with respect to a parameter and the derivative of the total Hamiltonian $\hat{H} = \hat{H}_0 + \hat{H}_{int}$ with respect to the same parameter \cite{LandauLifshitz5, Abrikosov1975}. 
Then one has:
\beq \label{force.derivative} \frac{\partial \Omega(\mb{r})}{\partial \mb{r}}=\bigg \langle\frac{\partial \hat{H}_{int}(\mb{r})}{\partial \mb{r}}\bigg \rangle
.
\eeq
\begin{figure}[t!]
    \centerline{
    \includegraphics [width=0.2\linewidth]{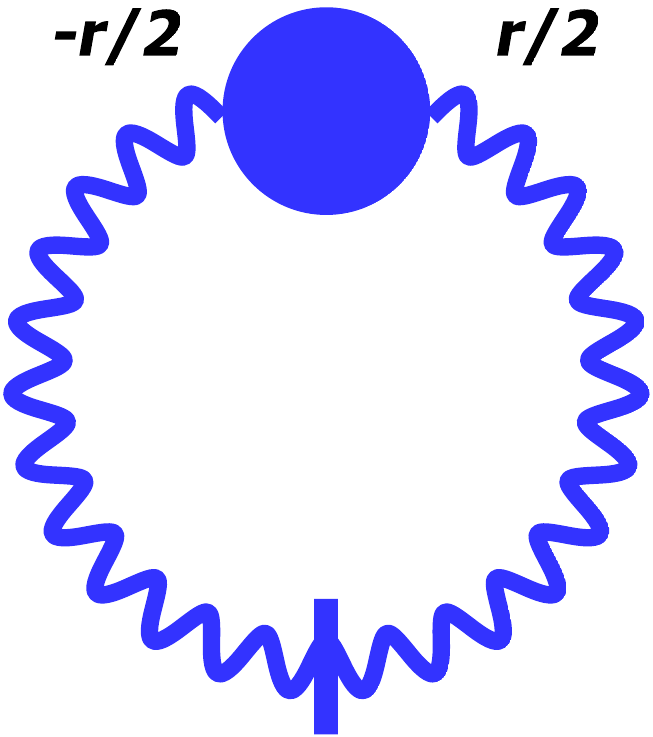}
}
    \caption{Diagrammatic representation of  the derivative of the thermodynamic potential with respect to distance of the impurities.}
    \label{fig:Derivative.sum}
\end{figure}

\begin{figure}[t!]
    \centerline{
        \includegraphics [width=1\linewidth]{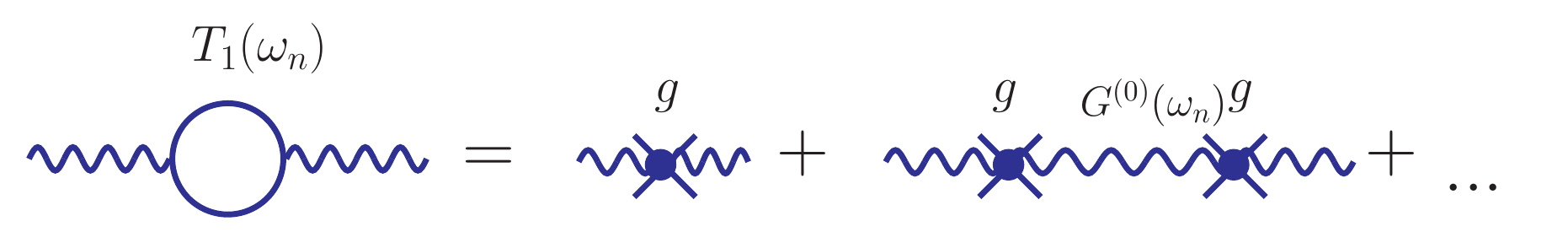} }
    \caption{Definition of the single-particle T-matrix $T_1(\omega_n)$.}
    \label{fig:Tmatr}
\end{figure}

\begin{figure}[t!]
    \centerline{
        \includegraphics [width=1\linewidth]{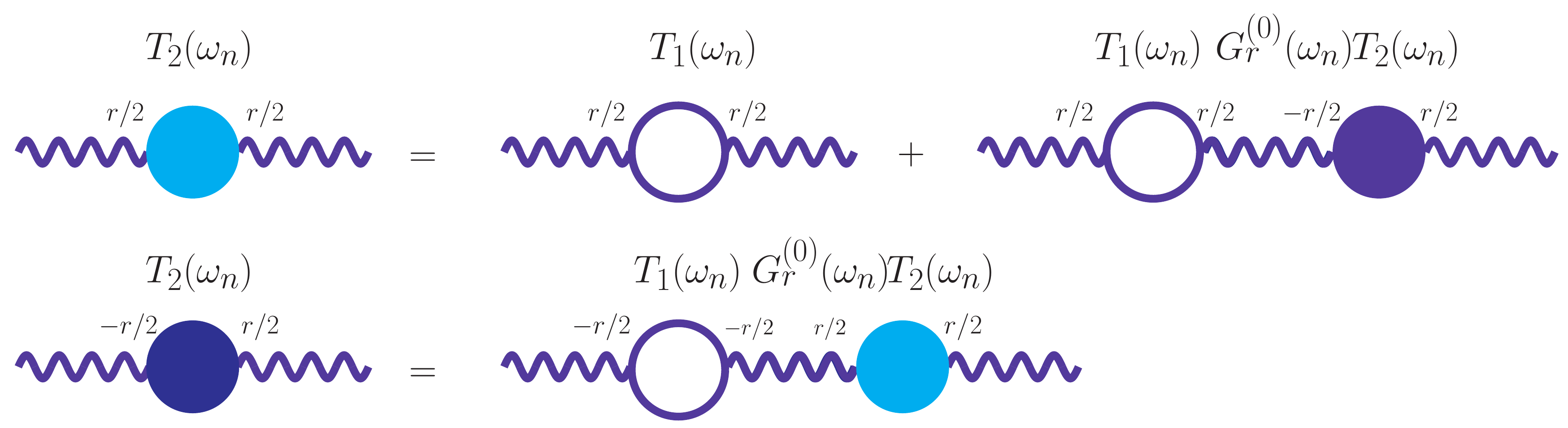}
    }
    \begin{minipage}{0.5\textwidth}\caption{
     Diagrammatic representation of the two-impurity $T$-matrix. The empty circles here correspond to $T_1(\omega_n)$, light blue and dark blue circles mean $T_2\left(\frac{\mb{r}}{2},\frac{\mb{r}}{2},\omega_n\right)$ and $T_2\left(-\frac{\mb{r}}{2},\frac{\mb{r}}{2},\omega_n\right)$ respectively, and the wavy lines are the Green's functions $G^{(0)}_r(\omega_n)$. }
     \label{fig:2imp_T-matrix} \end{minipage}
\end{figure}

The right side of Eq.(\ref{force.derivative}) can be found using the T-matrix approach. The corresponding diagram is presented in Fig. \ref{fig:Derivative.sum}.
The solid line with a tick stands for the derivative of the Green's function  $\frac{\partial G^{(0)}_r(\omega_n)}{\partial r}$. The circle is the two-impurity scattering $T$-matrix $T_2(-\frac{\mb{r}}{2},\frac{\mb{r}}{2},\omega_n)$.  
Then Eq.\eqref{force.derivative} takes the form: 
\beq \nn \frac{\partial\Omega(r)}{\partial r}=T\sum_n\frac{\partial G^{(0)}_r(\omega_n)}{\partial r}T_2\left(-\frac{\mb{r}}{2},\frac{\mb{r}}{2},\omega_n\right). \eeq
$T_2\left(-\frac{\mb{r}}{2},\frac{\mb{r}}{2},\omega_n\right)$ can be deduced from the single impurity scattering matrix $T_1(\omega_n)$. The series for $T_1(\omega_n)$ is shown in Fig.\ref{fig:Tmatr}. The explicit form is
\begin{equation}
	T_1(\omega_n)=\frac{g}{1-gG^{(0)}(\omega_n)}. 
	\label{eq.T1matrix}
\end{equation}
The two-impurity $T$-matrix, represented in Fig.\ref{fig:2imp_T-matrix}, is given by the set of the following equations   
\bea \nonumber
&&T_2\left(\frac{\mb{r}}{2},\frac{\mb{r}}{2},\omega_n\right)=
T_1(\omega_n)+T_1(\omega_n)G^{(0)}_r(\omega_n)T_2\left(-\frac{\mb{r}}{2},\frac{\mb{r}}{2},\omega_n\right), \\
&&T_2\left(-\frac{\mb{r}}{2},\frac{\mb{r}}{2},\omega_n\right)=T_1(\omega_n)G^{(0)}_r(\omega_n)T_2\left(\frac{\mb{r}}{2},\frac{\mb{r}}{2},\omega_n\right).
\label{eq.2imp.Tmatrix}
\eea
Solving  Eqs.(\ref{eq.2imp.Tmatrix}), one gets
\beq
\frac{\partial\Omega(\mb{r})}{\partial \mb{r}}=-T\sum_{n=-\infty}^{\infty}\frac{(T_1(\omega_n) G^{(0)}_{\mb{r}}(\omega_n))^2}{1-\big (T_1(\omega_n)G^{(0)}_{\mb{r}}(\omega_n) \big )^2}
    \frac{\partial_r G^{(0)}_{\mb{r}}(\omega_n)}{G^{(0)}_{\mb{r}}(\omega_n)}.
\label{eq.Casimir.force}
    \eeq
The Casimir interaction can be found by integration of Eq. (\ref{eq.Casimir.force}) with a condition $\Delta\Omega(r)\rightarrow 0$ for $r\rightarrow \infty$:
\begin{equation}
U_{Cas}(r)\equiv \Delta\Omega(r)=
%T\sum_n\frac{(T_1(\omega_n) G^{(0)}_{\mb{r}}(\omega_n))^2}{1-\big (T_1(\omega_n)G^{(0)}_{\mb{r}}(\omega_n) \big )^2}
%\frac{\partial_r G_{\mb{r}}(\omega_n)}{G^{(0)}_{\mb{r}}(\omega_n)}
T\sum_{\omega_n>0} \ln\left( 1-\big (T_1(\omega_n)G^{(0)}_{\mb{r}}(\omega_n) \big )^2\right).
\label{eq.thermodynamic.potential}
\end{equation}

Now, as we have established the general expression for the Casimir energy via the single particle scattering matrix and Green's functions, it is worth evaluating these Green's functions in various dimensions. The Green's function $G^{(0)}_{r}(\omega_n)$ ($r \neq 0$) can be explicitly calculated for the linear boson spectrum $\omega_{\mb{k}} = ck$ from Eq.(\ref{eq.GreenR}). It yields in the dimensions $D=1-3$ (see Appendix B for details)
\begin{equation}
 G^{(0)}_{r}(\omega_n)
=\begin{cases}
    -\frac{|\omega_n|}{2c}e^{-\frac{|\omega_n|}{c}r}, & D=1, \\
    -\frac{|\omega_n|^2}{2\pi c^2}K_0\left(\frac{|\omega_n|}{c}r\right), & D=2,\\
    -\frac{|\omega_n|^2}{4\pi r c^2}e^{-\frac{|\omega_n|}{c}r}, & D=3.
\end{cases}
\label{eq.GreenR.evaluated}
\end{equation}
where $K_0(x)$ is the modified Bessel function of the second kind.\\
We would like to note that 
the large distance scaling of $G^{(0)}_r(\omega_n)$ is universal and can be expressed in the form
\beq \label{eq.GreenR.high.dimensions}
G^{(0)}_r(\omega_n)\underset{r\frac{|\omega_n|}{c}\gg 1}{\sim} |\omega_n|^{\frac{D+1}{2}}r^{-\frac{D-1}{2}}e^{-\frac{|\omega_n|}{c} r}.
\eeq
Due to the exponential dependence of $G^{(0)}_r(\omega_n)$ on the energy, the leading contribution to the Casimir effect comes from low energies $\omega_n\ll\omega_n^*=c/r $ at large distances.\\
\subsection{Second order of the perturbation theory}
\begin{figure}[t!]
    \centerline{
        \includegraphics [width=0.5\linewidth]{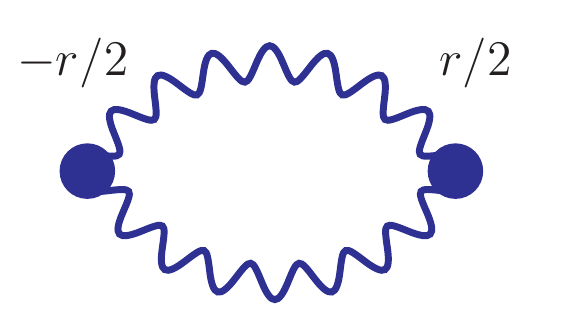} }
    \caption{$U_{Cas}(r)$ in the second order of the perturbation theory.}
    \label{fig:Second_derivative}
\end{figure}

The lowest order term in $g$ contributing to the Casimir interaction is of the second order, which is given by the diagram depicted in Fig. \ref{fig:Second_derivative}:
\beq \label{force_second} U_{Cas}(r)=\int_{0}^{\infty}\frac{d\omega_n}{2\pi}\left(gG^{(0)}_r(\omega_n)\right)^2. \eeq
 Using the expression for $G^{(0)}_r(\omega_n)$ given by Eq. \eqref{eq.GreenR.evaluated} and Eq. \eqref{eq.GreenR.high.dimensions}, we get that
 the Casimir potential in the second order of the perturbation theory obeys the law $U_{Cas}(r) \sim r^{-(2D+1)}$.
In $D=1..3$, we get the following expressions:
\beq \nn
U_{Cas}(r) = \begin{cases}
    -\frac{g^2c}{32\pi r^3}, & D=1, \\
    -\frac{27g^2c}{2048r^5}, & D=2,\\
    -\frac{3g^2c}{64\pi^2r^7}  , & D=3.
\end{cases}
\eeq
The expression for $D=1$ agrees with earlier results obtained in \cite{Schecter2014,Pavlov2018}.

\subsection{$T$-matrix approximation}

\paragraph{Continuum limit.} Keeping only the leading terms in the low energy expansion, one can write down the single impurity $T_1(\omega_n)$-matrix of the phonon-impurity scattering in the following form (for details see Appendix B):
\beq \label{eq.Tmatrix}
T_1(\omega_n)\simeq \begin{cases}
     \frac{1}{a^{-1}+A_1\frac{|\omega_n|}{c}}, & D=1, \\
      \frac{1}{a^{-1}+A_2\frac{|\omega_n|^2}{c^2}\ln|\frac{\omega_c}{\omega_n}|}, & D=2,\\
      \frac{1}{a^{-1}+A_3\frac{|\omega_n|^2}{c^2}} , & D\geq 3.
    \end{cases}
\eeq 
	The coefficients $a$, $A_D$ and $\omega_c$ can be determined in various transport experiments and can be considered phenomenological parameters.  Then  Eq. (11) directly relates the Casimir interaction to the physical properties of the phonon - single impurity scattering amplitude.  
 This parameter cannot be found in the frame of the considered microscopic theory, since it emerges as a consequence of the linear spectrum possessed by an effective low-energy theory.  
For numerical estimations of the Casimir forces we use further a mapping of the continuum model with linear dispersion on the lattice model presented in the next paragraph.

For the finite value of $a$, the $T_1$-matrix can be approximately set as $T_1\to a$ at small values of $\omega_n \ll \omega_n^*$. As one can see from Eq.\eqref{eq.Tmatrix}, the characteristic energy $\omega_n^*$ depends on the dimensionality and for $D=1-3$ reads:
 $\omega_{n,1D}^* = c /(a A_1)$, $\omega_{n,2D}^* = c \sqrt{1/(a A_2|\ln{\omega_c (\sqrt{a A_2}/c) }|)}$,
 $\omega_{n,3D}^* = c \sqrt{ 1/(a A_3)}  $.
 
Since $G^{(0)}_r(\omega_n)$ exponentially depends on $\omega_n r$, the leading contribution to the Casimir energy, as seen from Eq.\eqref{eq.Casimir.force}, comes from the energies $\omega_n \lesssim c/r$. It naturally defines the characteristic length $r_a\sim c/\omega_n^*$ of the change of the scaling behavior of the Casimir interaction.  At these energies, the $T_1(\omega_n)$ matrix can be approximated by the constant $a$. As a result, at large distances between the impurities $r\gg r_a$, the scaling of the Casimir interaction is the same as in the second order of the perturbation theory with the renormalized phonon-impurity coupling $U_{Cas}(r)\sim  T^2_1(\omega_n=0)/r^{(2D+1)}$. \\
For $g=g_{cr}$, $a\to\infty$ and $r_a\to\infty$. Hence, the energy dependence of  $T_1(\omega_n)$ matrix becomes important. The evaluation of Eq. \eqref{eq.Casimir.force} with $T_1(\omega_n)$ from Eq. \eqref{eq.Tmatrix} in the unitary limit shows that the Casimir interaction scaling in the leading order leads to
\beq \label{eq.UCas.unitary.cont}
U_{Cas}(r) \sim \begin{cases}
    \frac{1}{r}, & D=1, \\
    \frac{1}{r \ln^2 r}, & D=2,\\
    \frac{1}{r^{2D-1}}  , & D\geq 3.
\end{cases}
\eeq
 The analysis of the intermediate case of large but finite $r_a$ shows that at small distances $r\ll r_a$ the scaling in the leading approximation is described by  Eq. \eqref{eq.UCas.unitary.cont}.

\paragraph{Lattice model.}
Now we map the model on a lattice
in order to study the general properties of the $T$-matrix.
    We analyze an ideal harmonic cubic lattice described by
    $
    \nonumber H_0=\sum_i\frac{\hat{p}_i^2}{2m}+ \frac{m\omega^2_0}{2}\sum_{<i,j>}(\hat{u}_i-\hat{u}_j)^2,
    $
    with two embedded impurity atoms with a mass different from the mass of the atoms of the lattice.
    Here $\hat{p}_i$ and $\hat{u}_i$ are the momentum and coordinate operators, $m$ is the mass of the atoms of the cubic lattice and $m\omega_0^2$ is the potential term, $\omega_0=c/\delta$, and $\delta$ is the lattice constant. For simplicity, we put $\delta=1$.

 The excitations of the ideal harmonic lattice are  non-interacting phonons. The Hamiltonian reads $H_0=\sum_{\mb{k}}\omega_{\mb{k}}(b^{\dagger}_{\mb{k}}b_{\mb{k}}+\frac{1}{2})$.
    The  dispersion of the phonons on a lattice is given as $\omega_{\mb{k}}=c\sqrt{2D-2\sum_{\mb{\delta}}\cos\left(\mb{k\delta}\right)}$, where the summation is done over the nearest neighbors. The effect of the introduced impurity atoms can be considered a perturbation to the kinetic part of the Hamiltonian:  $V=\frac{g}{2m}(\hat{p}^2_a+\hat{p}^2_b)$, where $a$ and $b$ are impurity positions. The coupling constant $g =(1-m/M)$, with $m$ being the mass of the atom of the ideal lattice, and $M$ being the mass of the impurity atom. The momentum operator $\hat{p}_{\mb{k}}$ is quantized as $\hat{p}_{\mb{k}}=\textit{i}\sqrt{\frac{m\omega_{\mb{k}}}{2}}(\hat{b}^{\dagger}_{\mb{k}}-\hat{b}_{\mb{k}})$. This term $V$ is an equivalent of the phonon-impurity coupling given at Eq.(\ref{v1v2}). In terms of the bosonic operators $\hat{b}_{\mb{k}},\hat{b}^{\dagger}_{\mb{k}}$, it reads
\bea \nn V(\mb{r})&=&g\sum_{\mb{k},\mb{k'}}\sqrt{\omega_{\mb{k}}}\sqrt{\omega_{\mb{k'}}}(-\hat{b}^{\dagger}_{\mb{k}}\hat{b}_{\mb{k'}}\cos\frac{(\mb{k}-\mb{k'})\mb{r}}{2}\\
\nn &+&\hat{b}_{\mb{k}}\hat{b}_{\mb{k'}}\cos\frac{(\mb{k}+\mb{k'})\mb{r}}{2}+H.c.), \eea
$\mb{r}=\mb{r}_b-\mb{r}_a$. Rewriting this expression in terms of $\pi$ and $\bar{\pi}$, one obtains Eq.(\ref{v1v2}).
Then we define the Green's functions in accordance with Eqs. \eqref{eq.Green.function.definition}, \eqref{eq.GreenR}, and  \eqref{eq.Green0}.  
Integrations in $G^{(0)}_r(\omega_n)$ and $G(\omega_n)$ are performed over the Brillouin zone, and the integrals in Eqs. (\ref{eq.GreenR}) and (\ref{eq.Green0}) become finite:
\bea \nn G^{(0)}_r(\omega_n)&=&V^D_c\int_{BZ}\frac{d^Dk}{(2\pi)^D}\cos \mb{kr} \frac{\omega_{\mb{k}}^2}{\omega_n^2+\omega_{\mb{k}}^2},\\
\label{GreenLatt} G^{(0)}(\omega_n)&=&V^D_c\int_{BZ}\frac{d^Dk}{(2\pi)^D}\frac{\omega_{\mb{k}}^2}{\omega_n^2+\omega_{\mb{k}}^2}, \eea
$V_c^D$ is the elementary cell volume in $D$ dimensions.

Now we evaluate $G^{(0)}_r(\omega_n)$, $G^{(0)}(\omega_n)$, $T_1(\omega_n)$ and the Casimir interaction for the cubic lattice with the phonon spectrum $\omega_{\mathbf{k}}$:
\beq \label{deltaref} \omega_{\mathbf{k}}=2c\sqrt{\sum_{i=1}^D\sin^2\frac{k_i}{2}}, \eeq
where $D$ is the dimensionality.
The Green's functions  $G^{(0)}_r(\omega_n)$ and $G^{(0)}(\omega_n)$ are
\beq \nonumber G^{(0)}_r(\omega_n)= -f_D\left(\frac{|\omega_n|}{2c}\right), \eeq
\beq \nonumber G^{(0)}(\omega_n)= 1-f_D\left(\frac{|\omega_n|}{2c}\right), \eeq
where the function $f_D(x,r)$ does not contain any divergences and falls off exponentially with energy and distance. It can be analytically calculated in $D=1$ (see \cite{Pavlov2018}) and estimated in the leading approximation for higher dimensions.

The $T_1(\omega_n)$-matrix can be found exactly by summation of the contributing diagrams:
\beq
\label{Tmatrix.lattice.1d.exact} T_1(\omega_n)=\frac{1}{\frac{1-g}{g}+f_D\left(\frac{|\omega_n|}{2c},0\right)},
\eeq
The value $g=1$ corresponds to the unitary limit  $a = g/(1-g)\to\infty$. Namely, this limit corresponds to the scattering of bosons on a static impurity considered in \cite{Recati2005}. Far away from this limit, the low energy part of the $T$-matrix can be considered constant.

The function $f_{1D}$ is diven by Eq.(\ref{ffunc}), it reads:
$$
f_{1D}(x) =  \frac{x}{\sqrt{1+x^2}}(x+\sqrt{1+x^2})^{-2r}
$$
This expression in the limit of  small $\omega_n$  gives Eq.(\ref{eq.Tmatrix}), with $a=\frac{g}{1-g}$ and $A=\frac{1}{2c}$.

For the dimensions $D=2,3$, $a$ and $A$ are evaluated numerically.
For a cubic lattice in two and three dimensions, $G_r(\omega_n)$ in the leading approximation is identical to Eq.(\ref{eq.GreenR.evaluated}) in the limit of small Matsubara frequencies $\omega_n\ll c/\delta$. 
	The coefficients $a,\, A_D,\,\omega_c$ in the analytical expressions given above are fitted to match these numerical results. This gives us $G^{(0)}(\omega_n)=1-\frac{\omega_n^2}{2\pi c^2}\ln|\frac{\omega_c}{\omega_n}|$ in two dimensions. While the coefficient in front of the $\ln$ is universal in two dimensions, the value of $\omega_c$ depends on the parameters of the lattice. For the square lattice one can approximate $\omega_c^2\simeq 28$ in units of energy. The same calculation for the hexagonal lattice leads to the phonon spectrum $\omega^2_{\mb{k}}=\frac{8}{3}\big(\sin^2\frac{k_x}{2}+\sin^2\frac{k_x+\sqrt{3}k_y}{4}+\sin^2\frac{k_x-\sqrt{3}k_y}{4}\big)$,
    with $\omega_c^2 \approx 32$.
     In three dimensions,  $G^{(0)}(\omega_n)=1-\frac{1}{4 c^2}\omega_n^2$ in the leading approximation for low energies.

This allows us to approximate the Casimir interaction for the linearized spectrum as: 
\begin{widetext}
\beq \label{eq.Casimir1Dexpr} U_{Cas}^{(1D)}(r)=\int_0^{\infty}\frac{d\omega_n}{2\pi}\ln\left[1- \bigg(\frac{\frac{|\omega_n|}{2 c}e^{-\frac{|\omega_n|}{c}r}}{(1-g)/g+\frac{|\omega_n|}{2 c} }\bigg)^2\right], \eeq
\beq \label{Casimir2Dexpr} U_{Cas}^{(2D)}(r)=\int_0^{\infty}\frac{d\omega_n}{2\pi}\ln\left[1-\bigg( \frac{\frac{|\omega_n|^2}{2\pi c^2}K_0(\frac{|\omega_n|}{c}r)}{(1-g)/g+\frac{|\omega_n|^2}{2\pi c^2}\ln|\frac{\omega_c}{\omega_n}|}\bigg)^2\right], \eeq
\beq \label{eq.Casimir3Dexpr} U_{Cas}^{(3D)}(r)\simeq \int_0^{\infty}\frac{d\omega_n}{2\pi}\ln\left[1-\bigg(\frac{\frac{|\omega_n|^2}{4\pi r c^2}e^{-\frac{|\omega_n|}{c}r}}{(1-g)/g+\frac{|\omega_n|^2}{4c^2}}\bigg)^2\right], \eeq
\end{widetext}

The $r$ dependence of the Casimir interaction given by Eqs.\eqref{eq.Casimir1Dexpr}-\eqref{eq.Casimir3Dexpr} for various $g$ and $D=1-3$ is illustrated in Figs.\ref{fig:main} and \ref{fig:main1}. There are two regions, which are determined by the characteristic distance $r_a$. For $r\gg r_a$, one finds the universal scaling $U_{D}\sim \frac{1}{r^{2D+1}}$. At distances $r\lesssim r_a$ this interaction is not universal.   But at very short distances, $r\ll r_a =  c/\omega_n^*$, the Casimir interaction  can be approximated in the leading order as Eq. \eqref{eq.UCas.unitary.cont}.\\
In the unitary limit $g\rightarrow 1$ (the static limit in terms of the lattice model) due to the energy dependence of  $T_1(\omega_n)$, $U_{1D}(r)\sim 1/r$, $U_{2D}(r)\sim 1/{(r\ln^2 r)}$ and $U_{D}(r)\sim 1/r^{2D-3}$ for $D\geq3$ at any $r$ since $r_a\rightarrow \infty$.

\subsection{Casimir interaction at finite temperatures}
Finite temperatures affect the scaling of the Casimir interaction at large distances since the phonons get damping due to temperature. 
\begin{figure*}[t!]
\raggedleft
\begin{minipage}{.33\linewidth}
\subfloat[]{\label{main:a}\includegraphics[scale=.26,center]{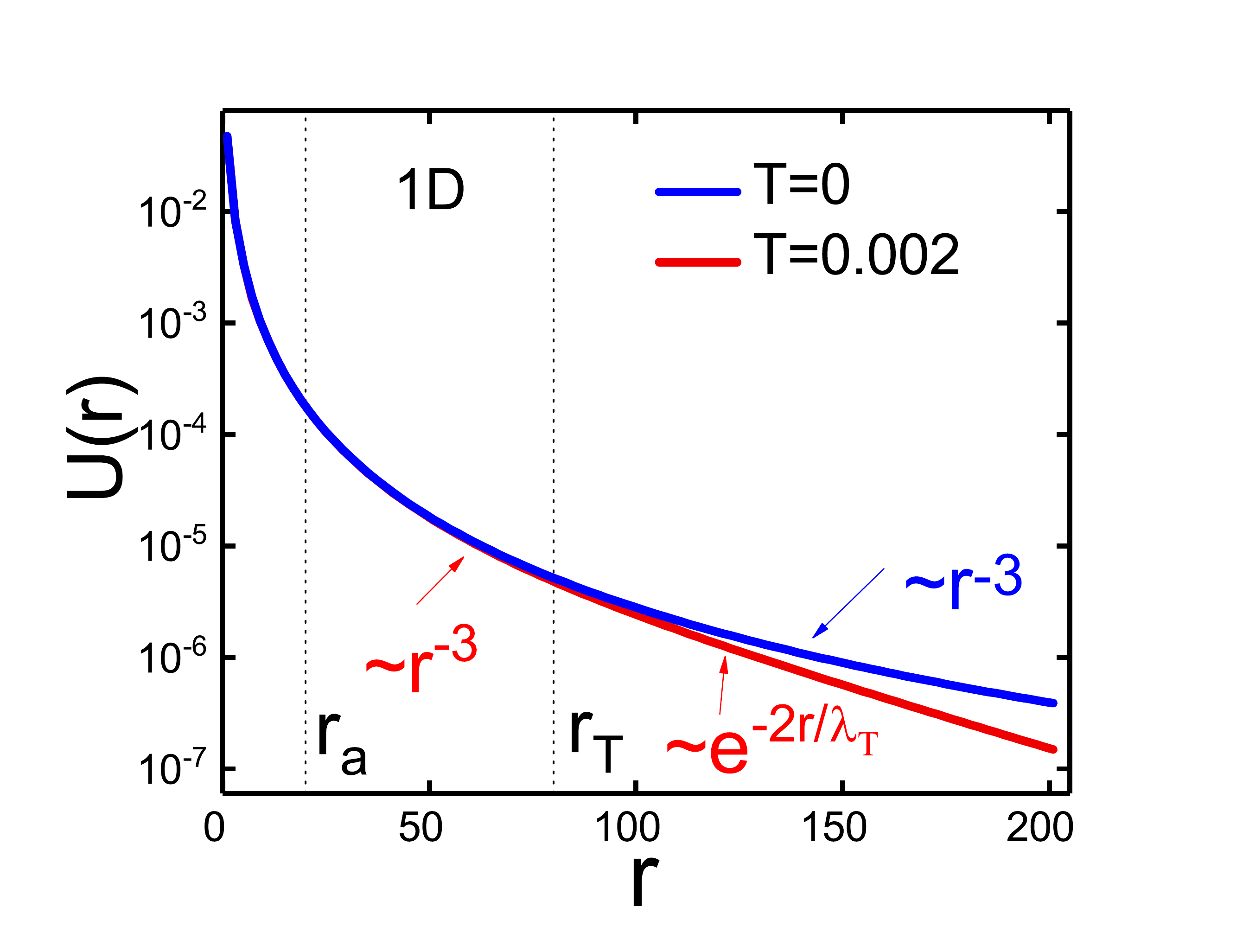}}
\end{minipage}%
\begin{minipage}{.33\linewidth}
\subfloat[]{\label{main:b}\includegraphics[scale=.26,center]{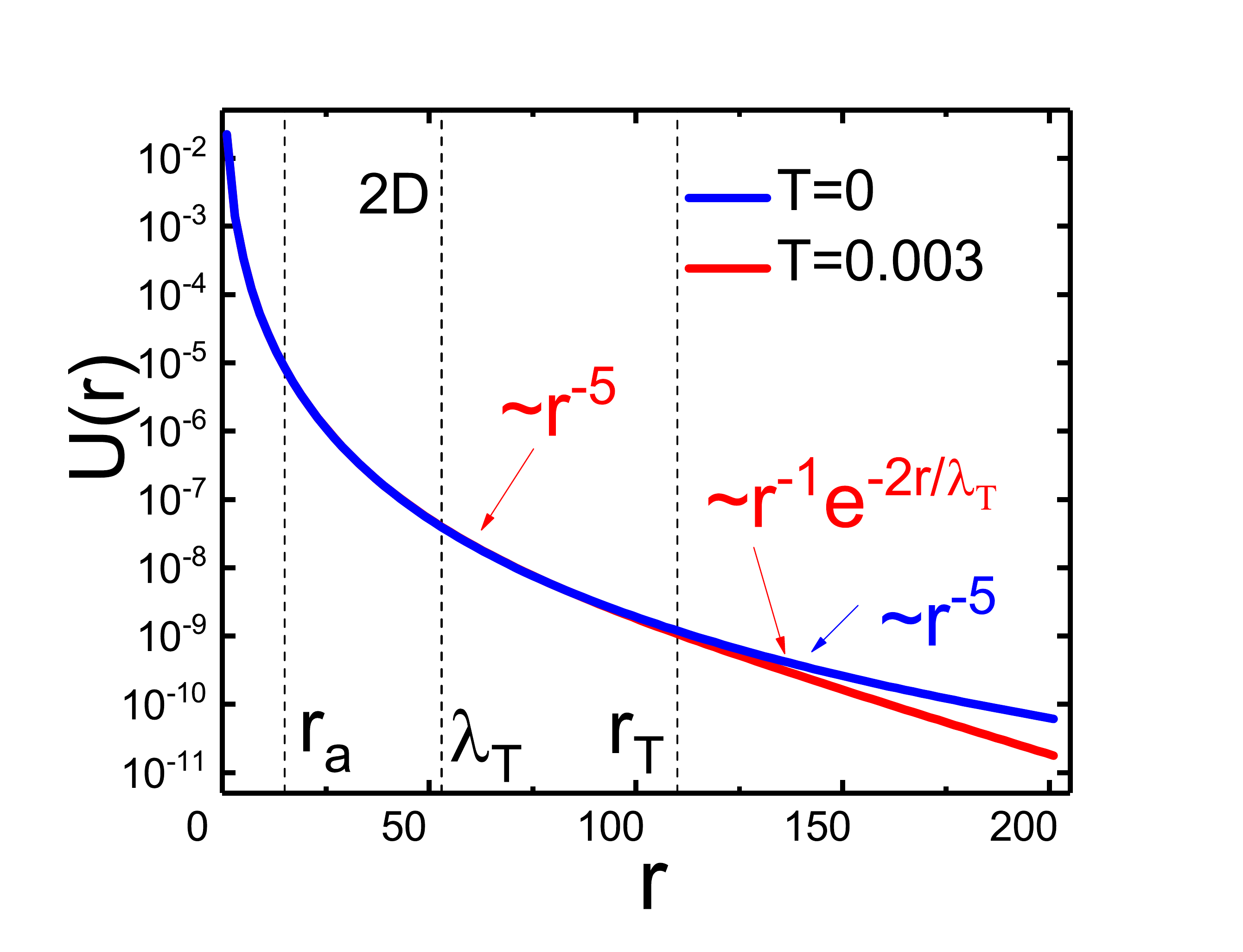}}
\end{minipage}
\begin{minipage}{.33\linewidth}
\subfloat[]{\label{main:c}\includegraphics[scale=.26,center]{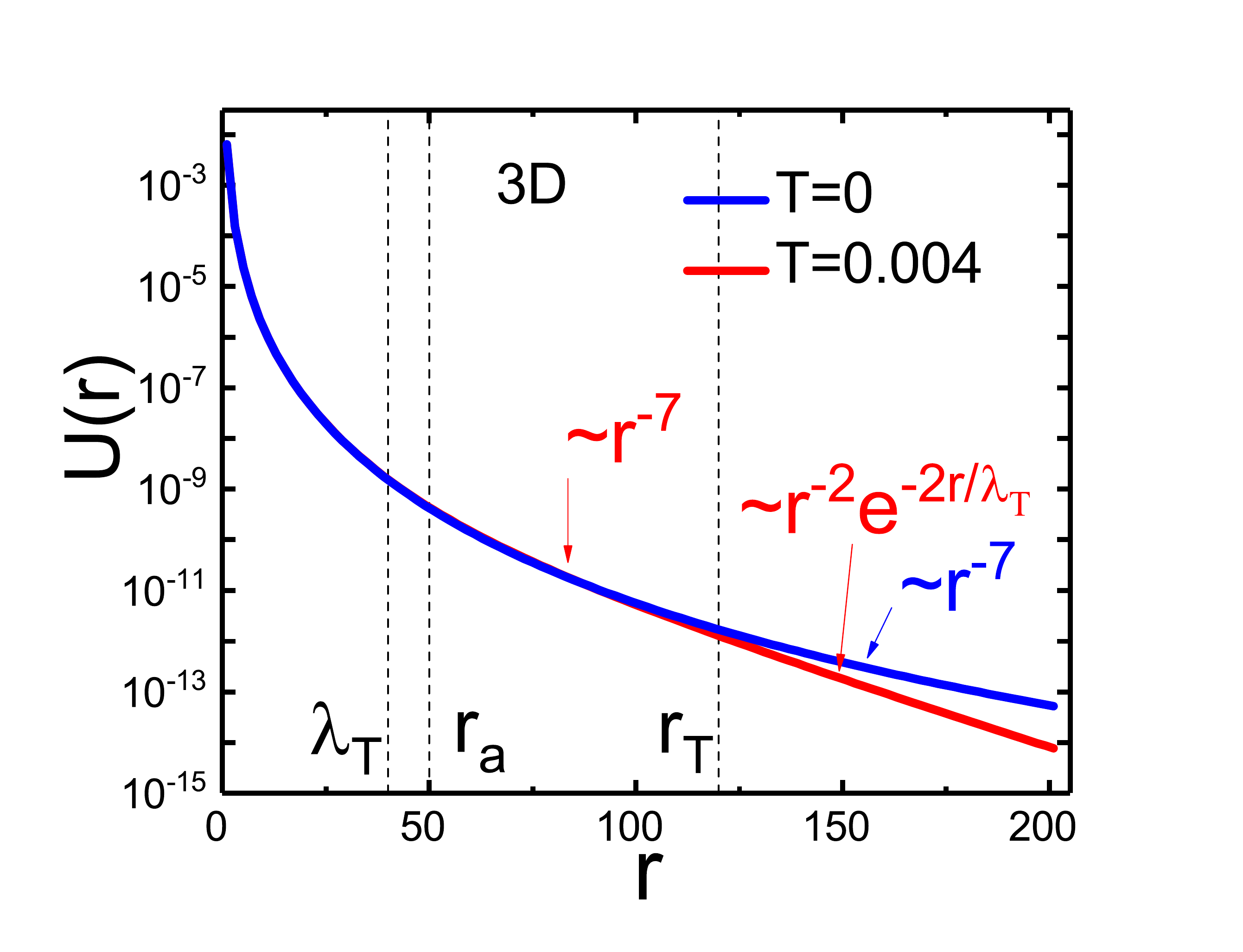}}
\end{minipage}
\caption{Casimir interaction of dynamic impurities at finite temperature and at $T=0$. $r_T$ is the effective length of the potential. In units of energy, (a) One-dimensional system with parameter $g=0.95$. Red line: $T=0.002\omega_0$. Blue line: $T=0$.  (b) Two-dimensional system with $g=0.99$. Red line: $T=0.003$. Blue line: $T=0$. $\lambda_T\simeq 53$ is the de Broglie wavelength, $r_T\simeq 110$.  (c) Three-dimensional system with $g=0.999$. Red line: $T=0.004$. Blue line: $T=0$. }
\label{fig:main}
\end{figure*}
The Casimir interaction at finite temperatures is given by Eq.\eqref{eq.thermodynamic.potential}. Since $G^{(0)}_r(\omega_n)\sim e^{-\omega_nr/c}$, there are two limiting cases of $r \gg \lambda_T$ and $r\ll \lambda_T$, where $\lambda_T = c/(2\pi T)$ is the thermal de Broglie wavelength.
In the former case, one restricts the summation by the first term. However in the latter case, one has to perform the summation over all Matsubara frequencies.

\begin{figure*}[t!]
\raggedleft
\begin{minipage}{.33\linewidth}
\subfloat[]{\label{main1:a}\includegraphics[scale=.26,center]{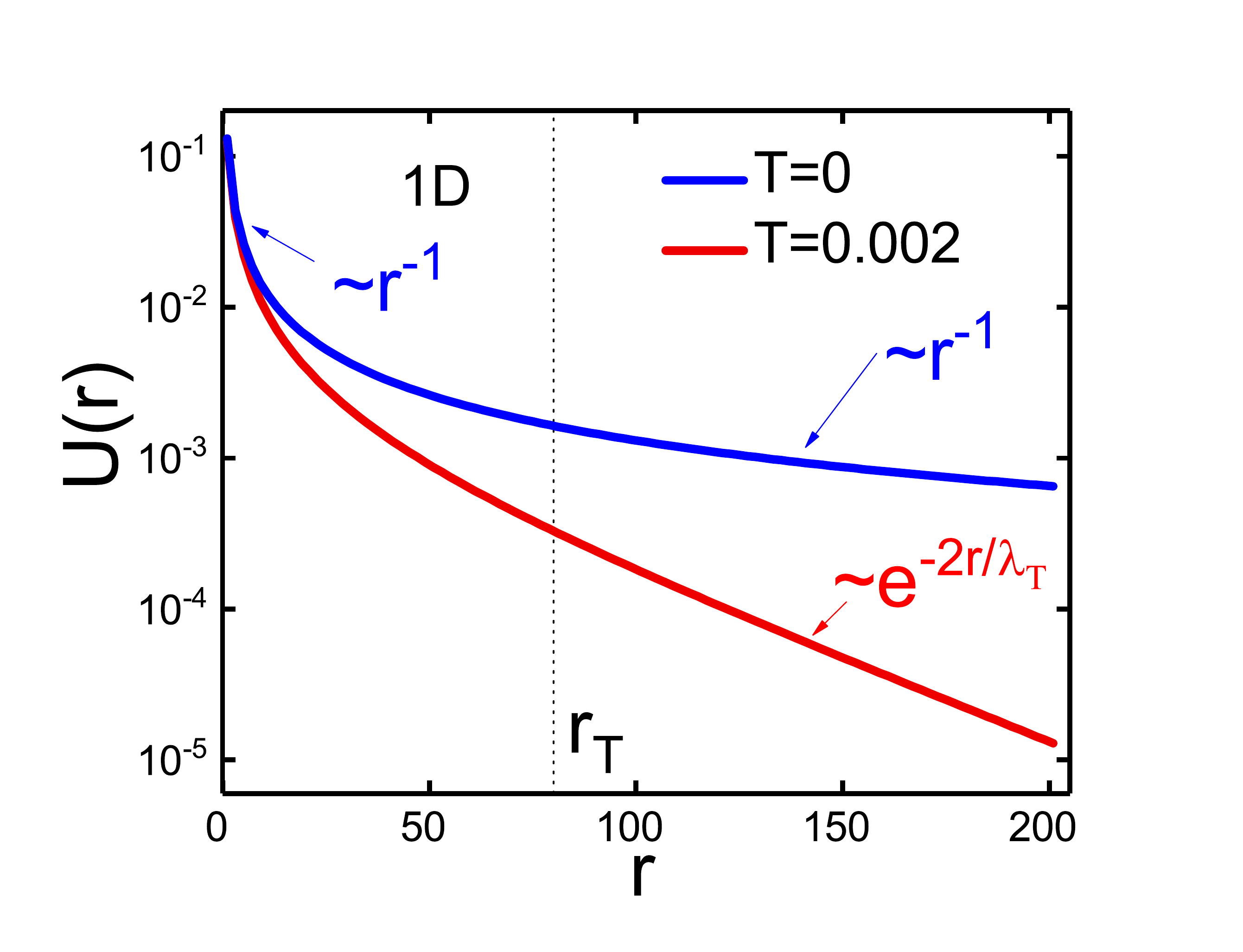}}
\end{minipage}%
\begin{minipage}{.33\linewidth}
\subfloat[]{\label{main1:b}\includegraphics[scale=.26,center]{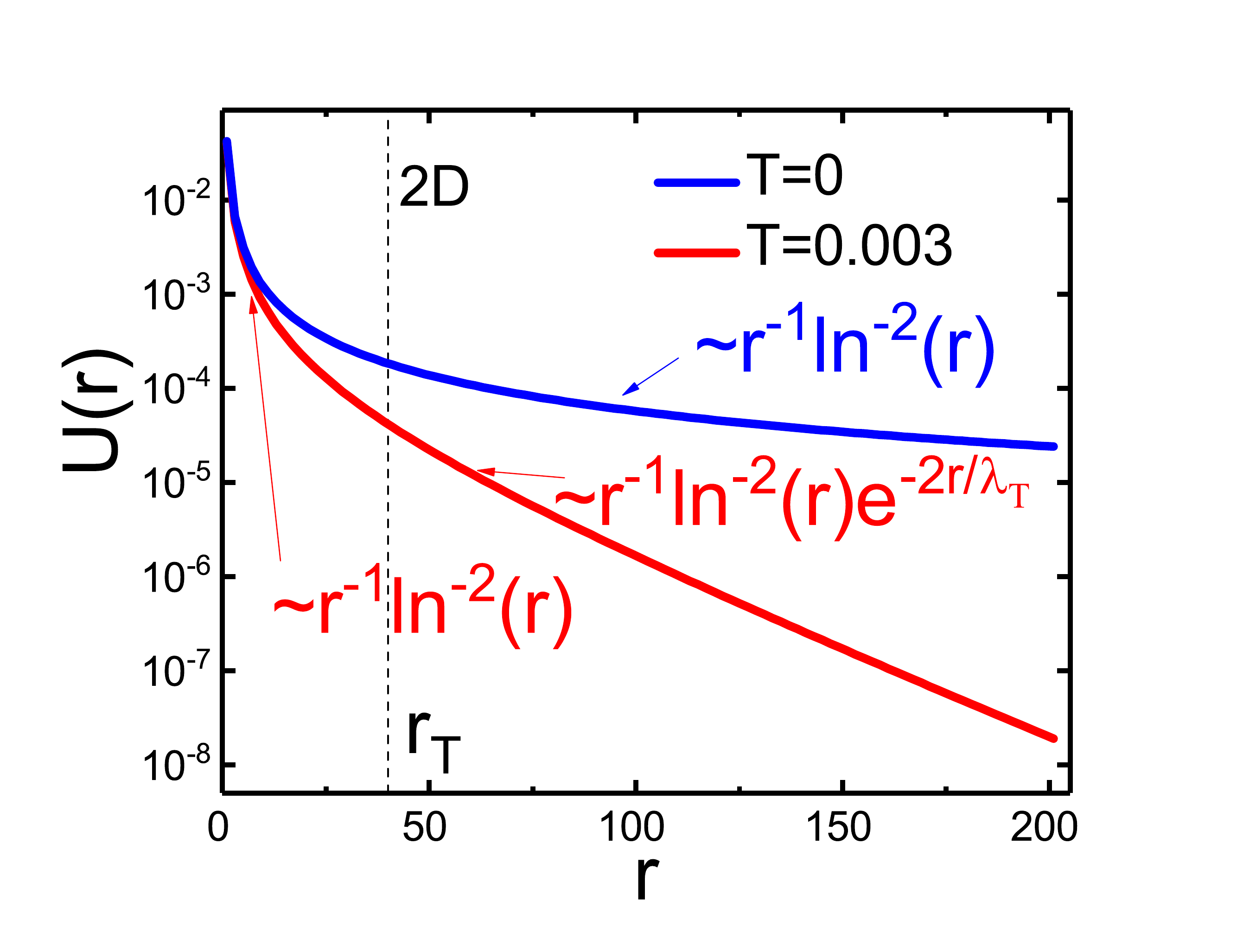}}
\end{minipage}
\begin{minipage}{.33\linewidth}
\subfloat[]{\label{main1:c}\includegraphics[scale=.26,center]{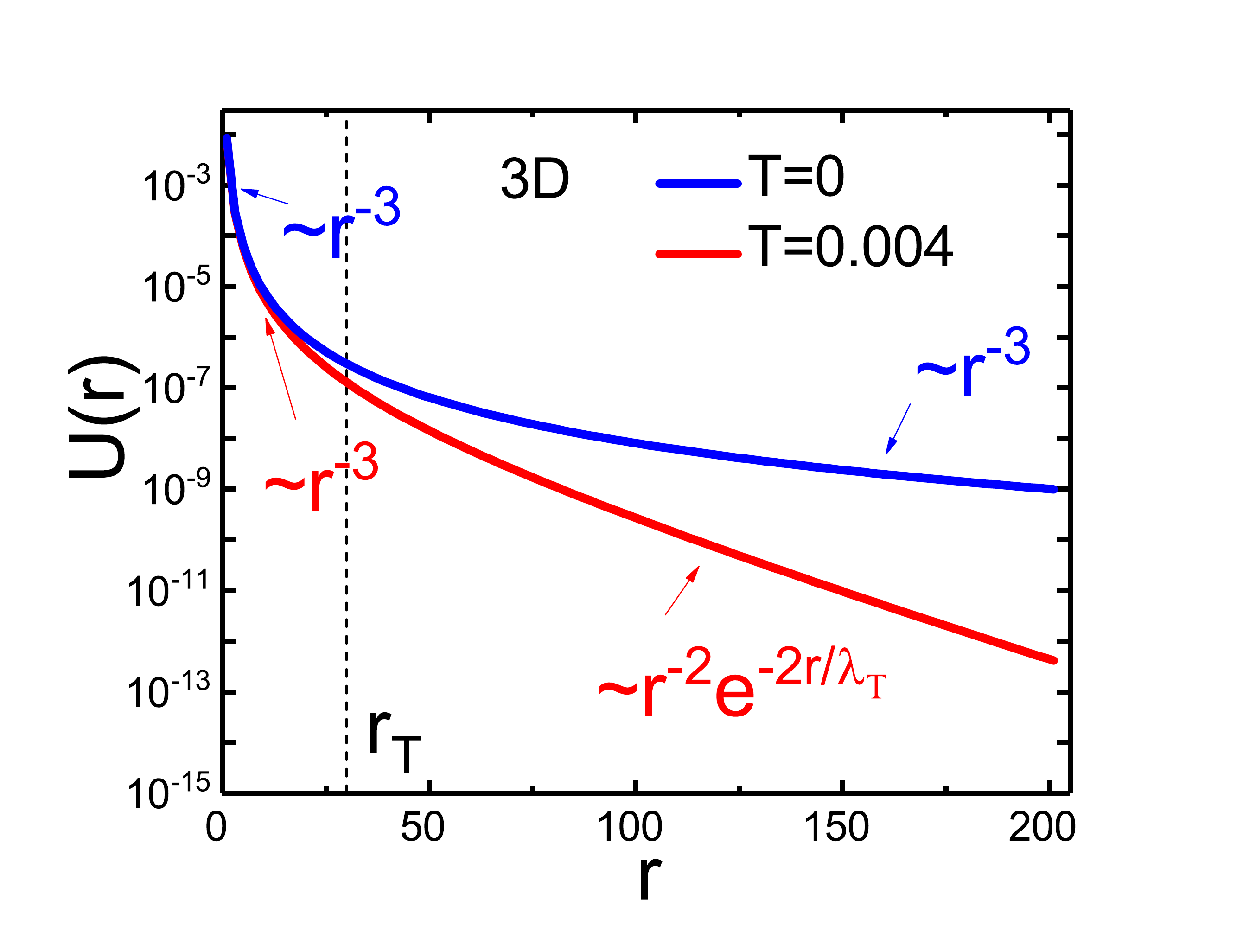}}
\end{minipage}
\caption{Casimir interaction of dynamic impurities  in the unitary limit ($g=1$). (a) One-dimensional system. Red line: $T=0.002$. Blue line: $T=0$; (b) Two-dimensional system. Red line: $T=0.003$. Blue line: $T=0$; (c) Three-dimensional system. Red line: $T=0.004$. Blue line: $T=0$.  }
\label{fig:main1}
\end{figure*}

To illustrate this point, it is worth considering a 1D system. In the explicit form, the thermodynamic potential reads:
\beq \nn U_{Cas}^{1D}(r)=T\sum_{n=1}^{\infty}\ln \left[ 1-\left(\frac{  e^{- n  r/\lambda_T}}{2\lambda_T (na)^{-1}+1}\right)^2 \right],
\eeq
where we use $a=g/(1-g)$. Now there are two characteristic lengths: the characteristic length $r_a$ and  $r_T$ of the order of the thermal de Broglie length $r_T\simeq \lambda_T$ [generally speaking, this parameter has a weak dependence on $g$ in $D>1$ and can differ from $\lambda_T$ by a numerical factor of the order of $1$, as illustrated in Fig. \ref{main:b}]. For $r\gg r_T$, the sum is dominated by the first Matsubara term and we get a universal exponential decay of the interaction:
\beq \label{eq.Casimir.1D.exprT} U_{Cas}^{1D}(r) \approx - T \frac{  e^{-2r/\lambda_T}}{\left(2\lambda_T /a +1 \right)^2} ,
\eeq
For $r \ll r_a, r_T$ the Casimir interaction follows the $r^{-1}$ law. For the intermediate distances $r_a \ll r \ll r_T$, the Casimir interaction falls off as $r^{-3}$.
A special consideration is required when we are exactly in the unitary limit $a\to \infty$. The decay of the Casimir interaction for $r\ll r_T$ in this case is precisely $r^{-1}$, and it further transfers to the exponential behavior Eq.\eqref{eq.Casimir.1D.exprT} for $r\gg\lambda_T$.
These dependencies can be seen in Figs. \ref{fig:main} and \ref{fig:main1}, in which the Casimir interaction at finite temperatures and at $T=0$ are presented. As a guideline, we depict approximate borders of the change of the Casimir law $r_a$ and $r_T$.
A detailed derivation of various limits is provided in Appendix C.

The effect of temperature in two-dimensional (2D) and three-dimensional (3D) systems can be calculated in the same way.  The calculations lead to: %
\be \nn U_{Cas}^{2D}=T\sum_{n=1}^{\infty}\ln \left[ 1-\frac{2( K_0(\frac{ nr}{\lambda_T}))^2}{\left(\frac{4\pi\lambda_T^2}{n^2a}+\ln\frac{\lambda^2_T\omega_c^2+(cn)^2}{(cn)^2}\right)^{2}} \right]
\ee
and
\beq \nn U_{Cas}^{3D}=T\sum_{n=1}^{\infty}\ln \left[ 1-\left(\frac{\frac{1}{\pi r}e^{-\frac{n r }{\lambda_T}}}{ \frac{4\lambda_T^2}{an^2}+1}\right)^2 \right]. \eeq
For $r\gg r_{T}$, the leading contribution to the thermodynamic potential comes from the first Matsubara frequency:
\be \label{2Das} U_{Cas}^{2D} \underset{r\gg r_{T}}{\simeq} -\frac{\frac{\pi T\lambda_T}{4r}  e^{-\frac{2 r}{\lambda_T}}}{\left(\frac{2\pi\lambda_T^2}{a}+\ln|\frac{\lambda_T\omega_c}{c}|\right)^{\!2}}\!
\ee
and
\beq \label{3Das} U_{Cas}^{3D}\underset{r\gg r_{T}}{\simeq}\frac{g^2T}{r^2\pi^2}\frac{e^{-\frac{2r}{\lambda_T}}}{(\frac{4\lambda^2_T}{a}+1)^2}. \eeq
The typical behavior is demonstrated in Figs. \ref{main:a} and \ref{main:b}. Note that the value of $r_T$ in two and three dimensions is different from the de Broglie wave length by some factor. This difference can be clearly seen in Fig. \ref{main:b}.   \\
One can see that at large distances $r\gg r_T$ the decay of the Casimir interaction is exponential in all dimensions. For  $r\ll r_T$ and $g<1$, there is a crossover from the short distance law to the $T=0$ long range law $r^{-(2D+1)}$.\\
The Casimir interaction in the unitary limit given in Fig.\ref{fig:main1} demonstrates the asymptotic behavior (\ref{eq.UCas.unitary.cont}) with temperature corrections similar to the ones given in Fig.\ref{fig:main} for the nonunitary case.

\section{Model for localized impurities in an external field}
 Now we consider the $\varphi\bar{\varphi}$ interaction. It corresponds to the external potential applied to two given atoms and trapping them at fixed positions. This situation was studied in \cite{Recati2005,Pavlov2018} for the one dimensional case. This case may be relevant for an experimental setup with trapped quantum gases proposed in \cite{Moritz2005, Recati2005}. It corresponds to the interaction for the field $\varphi(\mb{x})$ from Eq.(\ref{eq.H0}):
\bea \nn  \hat{H}_{int}=g\omega_0^2\Bigg( \left.\varphi\big (\mb{x})\bar{\varphi}\big(\mb{x}\big)\right\vert_{\mb{x}=-\frac{\mb{r}}{2}}+\left.\varphi\big (\mb{x})\bar{\varphi}\big(\mb{x}\big)\right\vert_{\mb{x}=\frac{\mb{r}}{2}}\Bigg),\\
\label{H.stat} \eea
The strength of the external potential is given by the value $g>0$, and $\omega_0$ is a unit of energy. \\
The Green's functions for this case are defined as
\beq \nn \tilde{G}^{(0)}(\mb{x},\mb{x}',\omega_n)=-\int_{0}^{\,\frac{1}{T}}d\tau e^{-\textit{i}\omega_n \tau}\langle T_{\tau}\big(\varphi(\mb{x},\tau)\bar{\varphi}(\mb{x}',0)\big) \rangle. \eeq
For the calculations, we need two Green's functions at the points $-\frac{\mb{r}}{2}, \frac{\mb{r}}{2}$, which we denote as
\bea \nonumber \tilde{G}^{(0)}_{r}(\omega_n)&\equiv &\tilde{G}^{(0)}\big(+\frac{\mb{r}}{2},-\frac{\mb{r}}{2},\omega_n\big)=\tilde{G}^{(0)}\big(-\frac{\mb{r}}{2},+\frac{\mb{r}}{2},\omega_n\big), \\
\nn \tilde{G}^{(0)}(\omega_n)&\equiv &\tilde{G}^{(0)}\big(+\frac{\mb{r}}{2},+\frac{\mb{r}}{2},\omega_n\big)=\tilde{G}^{(0)}\big(-\frac{\mb{r}}{2},-\frac{\mb{r}}{2},\omega_n\big).
 \eea
Their explicit expressions are 
\bea \nn \tilde{G}^{(0)}_{\mb{r}}(\omega_n)&=&\int\frac{d^D\mb{k}}{(2\pi)^D}\frac{\omega_0^2}{\omega_n^2+\omega^2_{\mb{k}}}e^{-\textit{i}\mb{kr}}, \\
\nn \tilde{G}^{(0)}(\omega_n)&=&\int\frac{d^D\mb{k}}{(2\pi)^D}\frac{\omega^2_0}{\omega_n^2+\omega^2_{\mb{k}}}.
 \eea
 \begin{figure*}[t!]
\raggedleft
\begin{minipage}{.33\linewidth}
\subfloat[]{\label{main.p:a}\includegraphics[scale=.26,center]{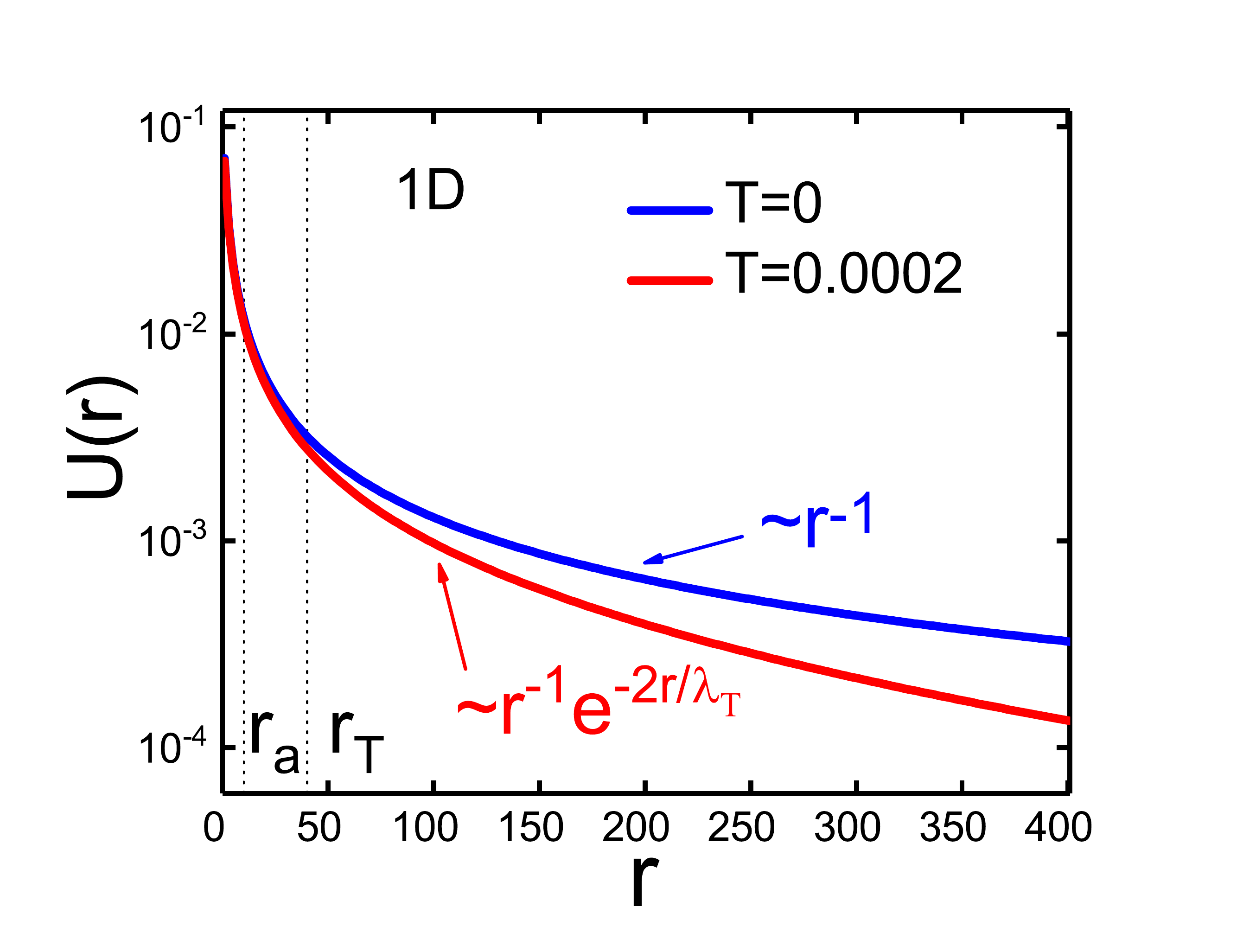}}
\end{minipage}%
\begin{minipage}{.33\linewidth}
\subfloat[]{\label{main.p:b}\includegraphics[scale=.26,center]{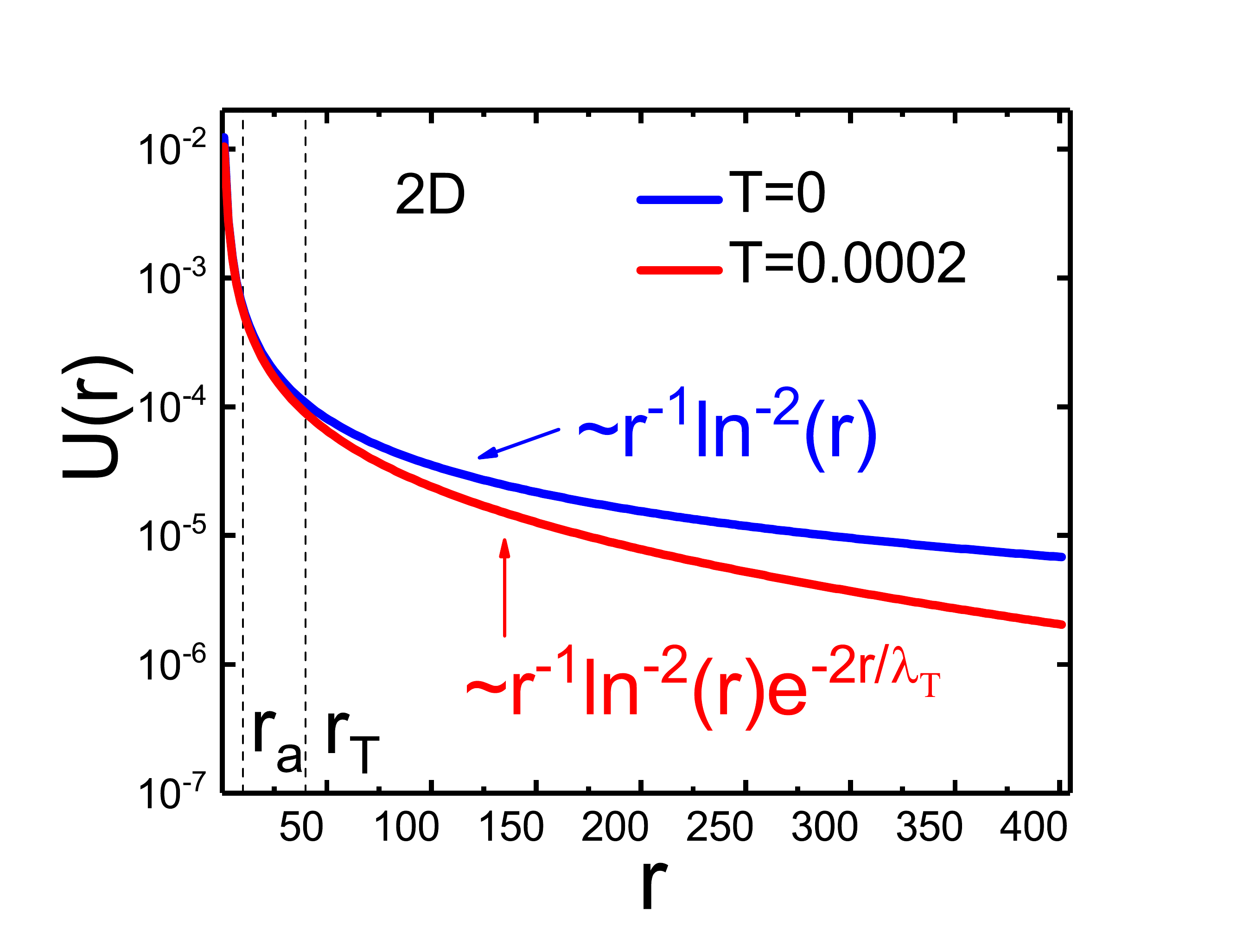}}
\end{minipage}
\begin{minipage}{.33\linewidth}
\subfloat[]{\label{main.p:c}\includegraphics[scale=.26,center]{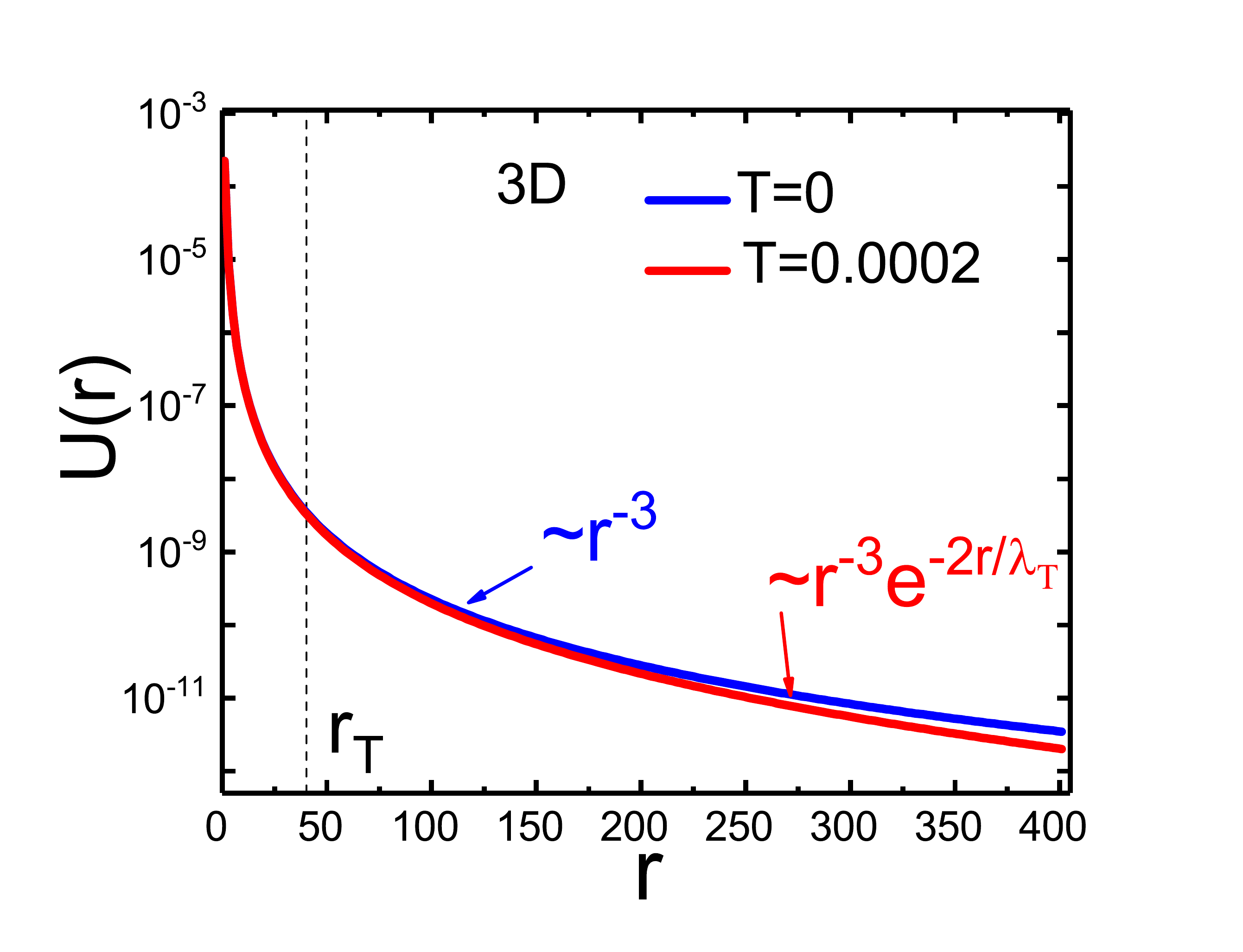}}
\end{minipage}
\caption{Casimir interaction for atoms trapped by external potential at a finite temperature and at $T=0$. $r_T$ is the effective length of the interaction, $g=2, \, \omega_0=1$. Blue line: $T=0$, red line: $T=0.0002$. (a) One-dimensional system. (b) Two-dimensional system.(c) Three-dimensional system. }
\label{fig:main.p}
\end{figure*}

The explicit form of $\tilde{G}^{(0)}_r(\omega_n)$ is given by the integration above and reads
\begin{equation}  \label{GrPot}
 \tilde{G}^{(0)}_{r}(\omega_n)
=\begin{cases}
    \frac{\omega_0^2}{2|\omega_n|c}e^{-\frac{|\omega_n|}{c}r}, & D=1, \\
    \frac{\omega_0^2}{2\pi c^2}K_0(\frac{|\omega_n|}{c}r), & D=2,\\
    \frac{\omega_0^2}{4\pi c^2r}e^{-\frac{|\omega_n|}{c}r}, & D=3.
\end{cases}
\end{equation}

The second order of the perturbation theory in $D=1$ for $T=0$ is given by 
\beq \nn \tilde U^{(2)}_{Cas}(r)=\int_0^{\infty}\frac{d\omega_n}{2\pi}\frac{g^2\omega_0^4e^{-2\frac{|\omega_n|}{c}r}}{4\omega_n^2c^2}.  \eeq
The integral diverges at the lower limit. This divergence can not be eliminated by introduction of the lattice model. The same infra-red divergence emerges in any order of the perturbation theory. 
However, the single impurity matrix $\tilde{T}_1(\omega_n)$, which is the sum of the diagrams in Fig.\ref{fig:Tmatr}, allows us to renormalize this divergence.\\
$\tilde{T}_1(\omega_n)$ is defined as
\beq \label{TTmatr} \tilde{T}_1(\omega_n)=\frac{g}{1-g\tilde{G}^{(0)}(\omega_n)}. \eeq
  
The explicit form of Eq.(\ref{TTmatr}) is:
\beq \label{TmPot} \tilde{T}_1(\omega_n)\simeq \begin{cases}
     \frac{-g}{1+\frac{g\omega_0^2}{|\omega_n |c}}, & D=1, \\
      \frac{-g}{1+\frac{g\omega_0^2}{2\pi c^2}\ln|\frac{\omega_0}{\omega_n}|}, & D=2,\\
      \frac{-g}{1+\frac{g\omega_0^2}{4c^2}} , & D= 3.
    \end{cases}
\eeq
Note that the $T$-matrix vanishes in the limit $\omega_n\to 0$ in $D=1,\, 2$. This means that the impurities become essentially transparent for long-range phonons. The leading contribution comes from phonons of the energy $c/r$. Therefore, the Casimir interaction cannot be approximated via scatterings at zero energy, in contrast to the model of dynamic impurities.      

Substituting Eq.(\ref{GrPot}) and Eq.(\ref{TmPot}) into Eq.(\ref{eq.thermodynamic.potential}), we calculate the Casimir interaction. It remains finite since $\tilde{T}_1(\omega_n)$ cancels the divergence of $\tilde{G}^{(0)}_r(\omega_n)$ at $\omega_n\to 0$.\\
The results are presented in Fig.{\ref{fig:main.p}}.
In the long-range limit $r\gg r_a$, the Casimir potential scales as
\beq \label{Upot} U_{Cas}(r)\sim
\begin{cases}
	\frac{1}{r}, & D=1,\\
	\frac{1}{r\ln^2 r}, & D=2,\\
	\frac{1}{r^3} , & D=3.
\end{cases}
\eeq
The characteristic length $r_a$ of the crossover to the long-range limit is smaller than for the dynamic impurities.
The difference stems from the low energy behavior of the $T$-matrix. In contrast to the dynamic impurities, the T-matrix goes quickly to the saturation point. The characteristic length $r_a$ is fully determined by the saturation energy $\omega_{n,D}^*$ of the $T$-matrix as $r_a\sim c/\omega_{n,D}^*$. The saturation energies are: $\omega_{n,1D}^*\sim g\omega_0^2/c$, $\omega_{n,2D}^*\sim \omega_0e^{-\frac{g\omega_0^2}{2\pi c^2}}$. Since $\tilde{T}_1\simeq const$ for $D=3$, $r_{a,3}=0$, and the $\frac{1}{r^3}$-law is fulfilled at all distances.

Comparing the long-range behavior with the unitary limit of the dynamic impurities, one finds that the scaling is exactly the same. It originates from the fact that both cases describe the same physical picture of two classical localized impurities of infinite mass. Mathematically, it follows from the identity (see Appendix D):

\beq \label{ident} T_1\left.(\omega_n)\right\vert_{g=1}G^{(0)}_r(\omega_n)=\tilde{T}_1\left.(\omega_n)\right\vert_{g=\infty}\tilde{G}^{(0)}_r(\omega_n). \eeq
Similar to the model of dynamic impurities, the power law changes to exponential at finite temperatures at long distances. The characteristic length $r_T$ is determined by the thermal de Broglie length with $r_T\sim c/2\pi T$. The crossover to the thermal regime is demonstrated in Fig. \ref{fig:main.p}.

\section{Discussion} 
Through the paper, we considered the Casimir interaction mediated by acoustic phonons. The phonon-mediated Casimir interaction between the dynamic impurities scales as $r^{-(2D+1)}$ in $D$ dimensions for a large distance in a continuum model. The result remains the same for large distances in a lattice model. 
A small difference can be found only at short distances $r\ll r_a$. The Casimir interaction at long distances is universal. The comparison of the Casimir interaction for the lattice model and the continuum model is given in Appendix E. The reason for the similarity of the results is the dominating contribution of low-energy phonons, for which the spectrum can be linearized.

It is worth discussing the relation between the considered model and the phonon mediated Casimir interaction in the Luttinger liquid in one dimension. 
The model, in contrast to the Luttinger liquid, does not contain vertex corrections to the phonon-impurity  scattering \cite{CastroNeto1996, Kane1992PRL, Kane1992, Schecter2014}. Therefore, the phonon-impurity scattering amplitude is treated phenomenologically in the low energy limit. The bare phonon-impurity interaction element coincides with that derived in the leading order in the Supplemental Material of \cite{Schecter2014}. 
 For 1D systems, our results for dynamic impurities in the second order of the perturbation theory are identical to the model of \cite{Schecter2014}. The results for static impurities are identical to the one considered in \cite{Recati2005}. Further, we demonstrated that the results for the two models converge in the limit of infinite mass of the impurities. \\
An impurity in a Luttinger liquid is dressed by a density depletion cloud \cite{Schecter2012}. These depletion clouds are relevant for interactions between static impurities in a Bose gas \cite{Recati2005, Petk} but are neglectable for dynamic impurities \cite{Schecter2014} and for static impurities in a Fermi gas \cite{Recati2005}. Our model corresponds to the two latter cases. \\
The model is restricted only on the phonon-mediated interaction. In real gases of cold atoms in addition to the phonon-mediated interaction, dipole-dipole interactions can arise (\cite{Derevianko1999}), but the phonon-induced Casimir interaction falls off much slower and therefore dominates over the dipole-dipole interactions (\cite{Pavlov2018}).

The phonon-induced Casimir interaction of impurities in cold atoms may be observable. As estimated in \cite{Schecter2014}, for the experimental setup \cite{Catani2012} of $^{40}K$ atoms with $^{87}Rb$ atoms as impurities, the Casimir interaction is expected to be $\sim 1kHz$ at $0.14\mu m$ separation between the impurities. For the parameters of this setup, we estimate $\omega_0\simeq 400kHz$ and the red line in Fig.\ref{main:a} corresponds to $T=40nK$, with $r_T\simeq 6\mu m$ (the interaction parameter $g$ used in our paper is the low-energy scattering amplitude given in (\cite{Schecter2014}).

An observable phonon-induced Casimir interaction may also arise in two and three dimensional lattices with phonons having a high Debye frequency. A particularly interesting material for this is solid hydrogen with deuterium impurities since in addition to a high Debye frequency, this system has a relatively large coupling parameter $g=\frac{1}{2}$ (due to the mass ratio $\frac{m_D}{m_H}=2$). With phonon energies of the order of $10meV$ \cite{Nielsen1973}, the Casimir energy between nearest neighbors impurities is expected to be $U_{Cas}\simeq 2\mu eV$. Other promising systems from this point of view are hydrates like  $H_3S$ \cite{Drozdov2015}, and superhydrides, such as $LaH_{10}$ \cite{Somayazulu2019}, etc. These materials, synthesized under high pressures, currently attract enormous attention due to unusually high critical temperatures of superconductivity, which is a consequence of high Debye frequencies. But considering multiatomic lattices is out of the scope of the present paper.\\

\section{Conclusion}
In conclusion, we have developed a theory of the phonon mediated Casimir interaction, based on the single-impurity scattering $T$-matrix. The model can be applied for impurities in quantum liquids in one dimensional systems, or lattices with embedded impurity atoms in two- and three dimensional systems.  The energy dependence of this quantity is vital for heavy impurities. We show that the energy dependence of the $T$-matrix determines the power law decrease of the Casimir interaction at short distances and at large distances in the unitary limit at $T = 0$. For the weak impurity scatterings, the Casimir interaction is universal at large distances. This T-matrix method is especially important for the consideration of the Casimir forces between two atoms in an external potential and allows us to obtain the nonperturbative results.

At finite temperatures, a new characteristic scale of the order of the thermal de Broglie wavelength appears.
For distances much larger than the de Broglie wave length, the Casimir interaction decays exponentially with the distance between the impurities.
Our results may be relevant to the proposed experimental setups of \citep{Moritz2005, Catani2012}

\paragraph*{Acknowledgments.} We have benefited from discussions with A. Kamenev, M. Kiselev, B. Altshuler, and I. Polishchuk. We thank U. Nitzsche for technical assistance. \\
The work was supported by the German Research Foundation (Deutsche Forschungsgemeinschaft) through the program DFG-Russia, BR4064/5-1. J.v.d.B is also supported by SFB 1143 of the Deutsche Forschungsgemeinschaft.

\bibliographystyle{apsrev}
\bibliography{casimir3}

\appendix
\section{Appendix A. The lattice model}\label{appA}
In this appendix, we derive the effective phonon-impurity interaction for a cubic harmonic lattice. The lattice is determined by the following Hamiltonian:
\begin{eqnarray} \nn  H=\sum_i\frac{\hat{\mathbf{p}}^2_i}{2m}+\frac{m\omega_0^2}{2}\sum_{|i-j|=1}(\hat{\mathbf{u}}_i-\hat{\mathbf{u}}_j)^2,
\end{eqnarray}
where $m$ is the mass of the atoms and $\omega_0$ is the characteristic energy of the nearest-neighbors interaction. 
After quantization and the Bogoliubov transformation the Hamiltonian takes the standard form for free phonons:
\begin{equation} 
\nn H=\sum_{\mb{k}}\omega_{\mb{k}}\left( b^{\dagger}_{\mb{k}}b_{\mb{k}}+\frac{1}{2} \right), 
\end{equation}
where $b^{\dagger}_{\mb{k}}$, $b_{\mb{k}}$ are creation and annihilation phonon operators and their dispersion in $D$ dimensions is 
\begin{eqnarray} \nonumber \omega_{\mathbf{k}}=\omega_0\sqrt{2D\nu_{\mathbf{k}}}, \,\nu_{\mathbf{k}}=1-\frac{1}{D}\sum_{i=1}^D\cos(q_i\delta),  \end{eqnarray}
 with $\delta$ being a lattice constant. For $q\delta\ll 1$ one gets  $\omega_k = c k $ with the sound velocity $c\omega_0\delta$.\\ 
Below, we consider two types of the perturbations at site  $i$:
\textit{(a)}  the mass of the atom at site $i$ is changed to $M$ without modification of the interaction with the nearest atoms (isotopic substitution),
\textit{(b)}  a local external harmonic potential is applied to the atom at position $i$. 

\paragraph{a)}
The perturbation is determined by the difference of the kinetic energy of the impurity atom at the site $i$ in comparison to the regular atom of the lattice: 
\begin{eqnarray} \nonumber \hat{H}_{int}=\frac{\hat{\mathbf{p}}_i^2}{2}\left( \frac{1}{M}-\frac{1}{m} \right). \end{eqnarray}
Introducing $g=\left(1-\frac{m}{M} \right)$, one can rewrite the expression in terms of phonon operators: 
\bea \nn H_{int}=-g\sum_{\mb{k},\mb{k'}} 
\sqrt{\omega_{\mathbf{k}} \omega_{\mathbf{k'}}}
\left(b_{\mb{k}}b^{\dagger}_{\mb{k'}}e^{-\textit{i}\mb{r}_i(\mb{k}-\mb{k'})}+b^{\dagger}_{\mb{k}}b_{\mb{k'}}e^{\textit{i}\mb{r}_i(\mb{k}-\mb{k'})}\right.\\
\nn -\left.b_{\mb{k}}b_{\mb{k'}}e^{-\textit{i}\mb{r}_i(\mb{k}+\mb{k'})}-b^{\dagger}_{\mb{k}}b^{\dagger}_{\mb{k'}}e^{\textit{i}\mb{r}_i(\mb{k}+\mb{k'})} \right),   \eea
which is in the short form: 
\beq \nn H_{int}=-g\pi(\mb{r}_i)\bar{\pi}(\mb{r}_i). \eeq

\textit{b)} The perturbation is 
\beq \nn H_{int}=gm\omega_0^2\left(\hat{\mathbf{u}}_i^2\right). \eeq
The expression in terms of phonon operators
\bea \nn H_{int}=g\sum_{\mb{k},\mb{k'}} \frac{1}{
\sqrt{\omega_{\mathbf{k}} \omega_{\mathbf{k'}}}}
\left(b_{\mb{k}}b^{\dagger}_{\mb{k'}}e^{-\textit{i}\mb{r}_i(\mb{k}-\mb{k'})}+b^{\dagger}_{\mb{k}}b_{\mb{k'}}e^{\textit{i}\mb{r}_i(\mb{k}-\mb{k'})}\right.\\
\nn +\left.b_{\mb{k}}b_{\mb{k'}}e^{-\textit{i}\mb{r}_i(\mb{k}+\mb{k'})}+b^{\dagger}_{\mb{k}}b^{\dagger}_{\mb{k'}}e^{\textit{i}\mb{r}_i(\mb{k}+\mb{k'})} \right),   \eea
which is in the short form: 
\beq \nn H_{int}=g\omega_0^2\varphi(\mb{r}_i)\bar{\varphi}(\mb{r}_i). \eeq

\section{Appendix B. Evaluation of the Green's function $G^{(0)}_r(\omega_n)$, $G^{(0)}(\omega_n)$ and the single particle $T-$matrix $T_1(\omega_n)$}\label{appB}
\paragraph*{Linear spectrum.} Here we evaluate Eq.(\ref{eq.GreenR})  for the linear spectrum $\omega_n=c|\mb{k}|$ at large distances.  For $r\neq 0$, it reads
\beq \nn G^{(0)}_r(\omega_n)=-\int\frac{dk_{\parallel}d^{D-1}\mb{k}_{\perp}}{(2\pi)^D}\frac{\omega_n^2}{\omega_n^2+c^2k_{\perp}^2+c^2k_{\parallel}^2}e^{-\textit{i}k_{\parallel}r}, \eeq
where dimensions is D.
We represented the vector $\mb{k}$ as $\mb{k}=\mb{k}_\perp+\mb{k}_{\parallel}$ and chose $\mb{k}_{\parallel}$ along $\mb{r}$. After integration over $k_{\parallel}$, we have
\beq \nn -C_D\left(\frac{\omega_n}{c}\right)^2\int_0^{\infty}dk_{\perp}\frac{k_{\perp}^{D-2}}{\sqrt{(\frac{\omega_n}{c})^2+k_{\perp}^2}}e^{-\sqrt{(\frac{\omega_n}{c})^2+k^2_{\perp}}r}, \eeq
where $C_D=\frac{\pi}{(2\pi)^D}\int d\Omega_{D-1}$ is a constant containing all angular integrations.\\
We renomalize the momentum introducing a new dimensionless variable $q$, defined as $k_{\perp}\rightarrow q\frac{\omega_n}{c}$. The contribution of large values of the momentum to the integral is exponentially small, so only small momenta matter here. It turns the integral into
\bea \nn &-&C_D\left(\frac{|\omega_n|}{c}\right)^D\int_0^{\infty}dq\frac{q^{D-2}}{1+q^2}e^{-\sqrt{1+q^2}\frac{|\omega_n|}{c}r}\underset{r\frac{|\omega_n|}{c}\gg 1}{\simeq }\\
\nn &-&C_D\left(\frac{|\omega_n|}{c}\right)^De^{-\frac{|\omega_n|}{c}r}\int_0^{\infty}q^{D-2}e^{-\frac{q^2}{2}\frac{\omega_n}{c}r}dq. \eea
The remaining integral can be evaluated exactly and gives $2^{\frac{D-3}{2}}(\frac{|\omega_n|}{c}r)^{-\frac{D-1}{2}}\Gamma(\frac{D-1}{2})$. It leads us to Eq.(\ref{eq.GreenR.high.dimensions}):
\beq \nn G^{(0)}_r(\omega_n)\simeq C_D2^{\frac{D-3}{2}}\Gamma \left(\frac{D-1}{2}\right)\left(\frac{|\omega_n|}{c}\right)^{\frac{D+1}{2}}r^{-\frac{D-1}{2}}e^{-\frac{|\omega_n|}{c}r}. \eeq
$G^{(0)}(\omega_n)$ diverges on the upper limit. Therefore, a cutoff $\omega_c$ is introduced. This cutoff is used here formally, all the divergent terms are included in the T-matrix. Its value for impurity-phonon scatterings can be measured experimentally; and therefore, there are no real divergences in this approach. Then the integration yields
\beq \nn
G^{(0)}(\omega_n) \simeq \begin{cases}
    \frac{\omega_c}{\pi c}-\frac{|\omega_n|}{2c}, & D=1, \\
    \frac{\omega_c^2}{4\pi c^2}-\frac{|\omega_n|^2}{2\pi c^2}\ln|\frac{\omega_c}{\omega_n}|, & D=2,\\
    \frac{\omega_c^3}{6\pi^2 c^3}-\frac{\omega_c|\omega_n|^2}{2\pi^2 c^3}  , & D= 3.
\end{cases} \eeq
Since the $T_1(\omega_n)$-matrix is given by the diagrams in Fig.\ref{fig:Tmatr}, it has the form 
\beq \label{DTmatr} T_1(\omega_n)=\frac{g}{1-gG^{(0)}(\omega_n)}.\eeq
Substitution of $G^{(0)}(\omega_n)$ gives in the low energy limit the form of the $T$-matrix given in Eq. (\ref{eq.Tmatrix}).

\paragraph*{Lattice.} The integrals (\ref{eq.GreenR}) and (\ref{eq.Green0})  are convergent on the lattice:
\bea \nn  G^{(0)}_x(\omega_n)=V_c\int_{BZ}\frac{d^Dk}{(2\pi)^D}\frac{\omega^2_{\mb{k}}}{\omega_n^2+\omega^2_{\mb{k}}}e^{\textit{i}\mb{k}\cdot\mb{x}}=V_c\int_{BZ}\frac{d^Dk}{(2\pi)^D}e^{\textit{i}\mb{k}\cdot\mb{x}}  \\
\nn -V_c\int_{BZ}\frac{d^Dk}{(2\pi)^D}\frac{\omega_n^2}{\omega_n^2+\omega^2_{\mb{k}}}e^{\textit{i}\mb{k}\cdot\mb{x}}=\delta_{x,0}-f_D\left(\frac{|\omega_n|}{2c},x\right). \eea

 For D=1, on a square lattice the equation above is written as $G^{(0)}(\omega_n)=1-f_1(\frac{|\omega_n|}{2c},0)$, with $f_1(x,r)=\frac{x}{\sqrt{1+x^2}}(x+\sqrt{1+x^2})^{-2r}$. In the small-$\omega_n$ limit, this function turns into $G^{(0)}(\omega_n)=1-\frac{|\omega_n|}{2c}$. For higher dimensions, the structure $1-f_D(\frac{|\omega_n|}{2c},0)$, with some finite function $f_D$, remains.\\
For $D\geq 3$, the Green's function $G^{(0)}(\omega_n)$ can be approximated:
\beq \nn \scalebox{0.93}{$G^{(0)}(\omega_n)=1-\omega_n^2 V_c\int_{BZ}\frac{d^Dk}{(2\pi)^D}\frac{1}{\omega_n^2+\omega^2_{\mb{k}}}\simeq 1-\omega_n^2V_c\int_{BZ}\frac{d^Dk}{(2\pi)^D}\frac{1}{\omega^2_{\mb{k}}}$}. \eeq
This means, that $G^{(0)}(\omega_n)\simeq 1-A_D\omega_n^2$, with  constant $A_D$.\\
For $D=1$, one gets $ G^{(0)}(\omega_n)-1\sim \omega_n$.\\
For $D=2$ the integral is $G^{(0)}(\omega_n)-1\sim \omega_n^2\ln|\omega_n|$.\\
\begin{figure*}[t!!]
\raggedleft
\begin{minipage}{.45\linewidth}
\subfloat[]{\label{relation:a}\includegraphics[scale=.35,center]{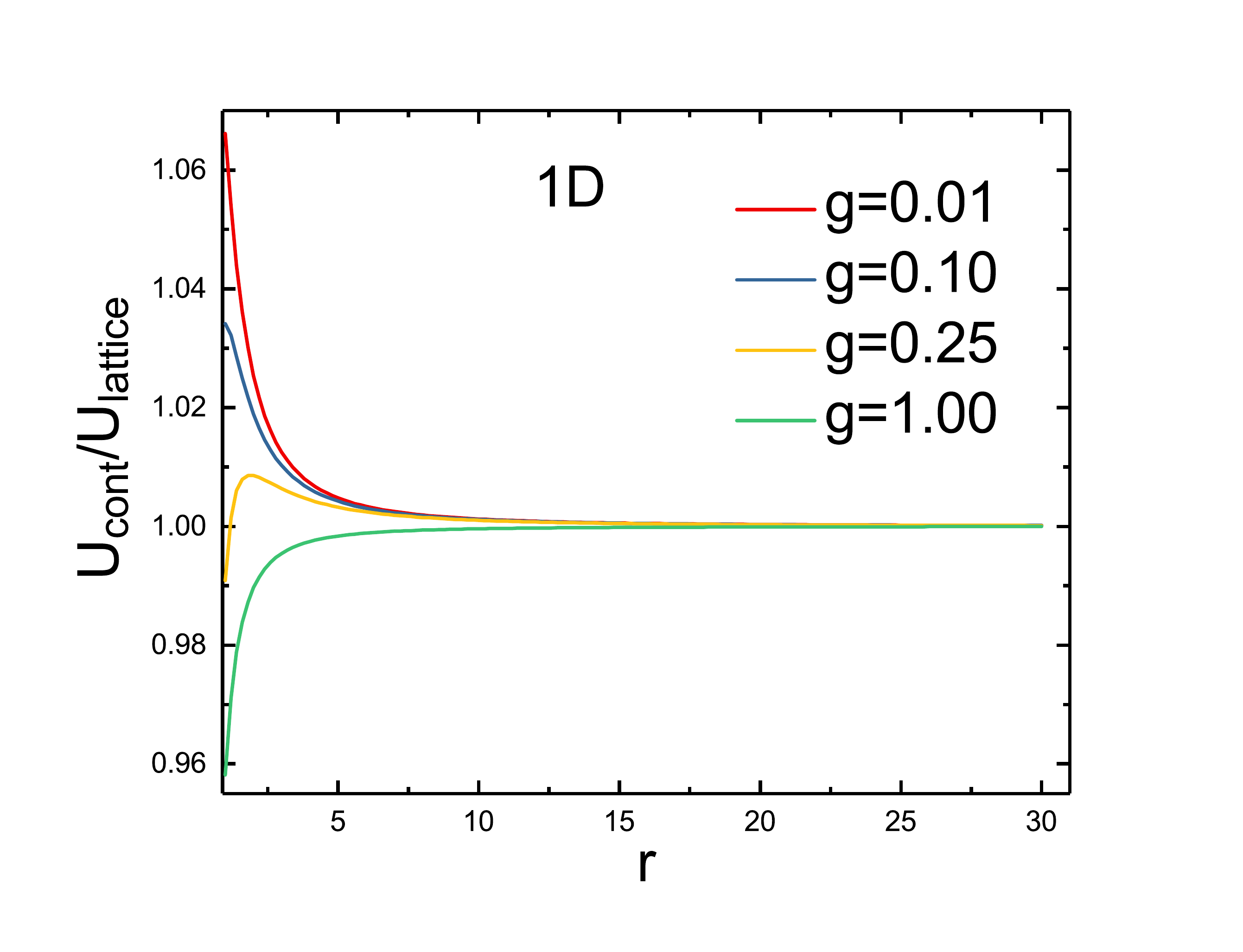}}
\end{minipage}
\begin{minipage}{.45\linewidth}
\subfloat[]{\label{relation:b}\includegraphics[scale=.35,center]{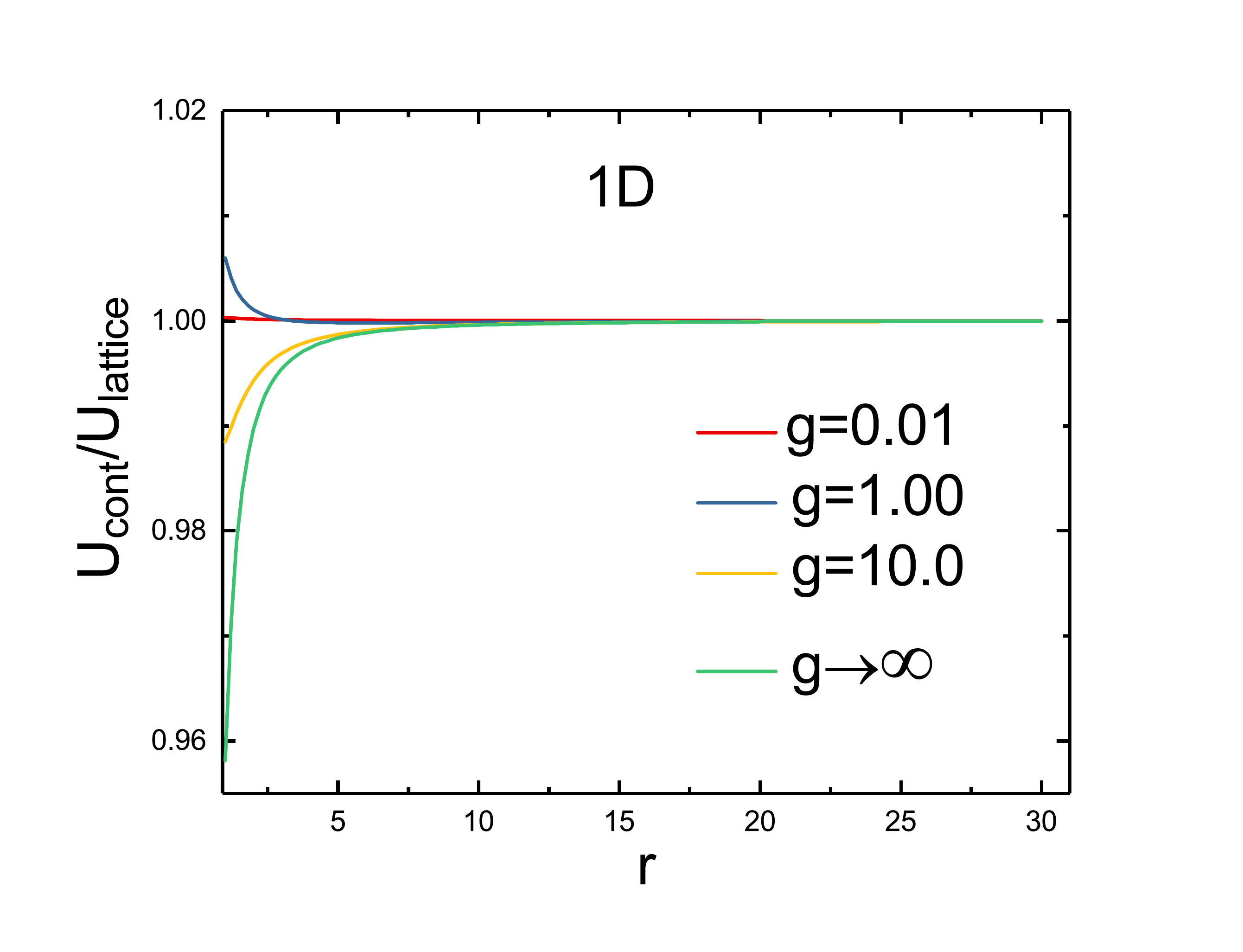}}
\end{minipage}
\caption{The ratio of the 1D Casimir interaction in the continuum limit $U_{cont}$ and the lattice model$U_{lattice}$. Here T=0. a) Dynamic impurities. Red line - $g=0.01$; blue line - $g=0.1$; yellow line - $g=0.25$; green line - $g=1$. b) Impurities localized in the external potential. Red line - $g=0.01$; blue line - $g=1$; yellow line - $g=10$; green line - $g\rightarrow \infty$.}
\label{fig:relation}
\end{figure*}

\section{Appendix C. Asymptotic behavior of the Casimir interaction at finite temperatures}\label{appC}
\paragraph*{Low temperature corrections in one dimension.} At low temperatures and short distances, when $r\ll \lambda_{T}$, we use the Euler-Maclaurin formula up to the first term to approximate the sum:
\bea \nonumber U_{Cas}^{1D}(r)&\underset{r\ll \lambda_{T}}{\simeq}&c\int_{0}^{\infty}\frac{dx}{2\pi r}\ln\left[ 1-\left(\frac{\frac{gx}{2r}e^{-x}}{1-g+\frac{g x}{2r}}\right)^2\right]\\
\nonumber &-&\left.\frac{1}{12}\frac{d}{dn}T\ln\left[1-\left(\frac{\frac{g\pi nT}{c}e^{-2\pi^2n\frac{ r }{\lambda_{T}}}}{1-g+\frac{g\pi n T}{c}} \right)^2 \right]^2\right \vert_{n=1}. \eea
At $g=1$ it gives
%\label{g1asympt}
\beq \nn U_{Cas}^{1D}(r)\underset{r\ll \lambda_{T}}{\simeq}-\frac{\pi c}{24 r}+\frac{2\pi^2 rT}{\lambda_{T}}, \eeq
while for small $g$ it gives
%\label{g01asympt}
\beq \nn U_{Cas}^{1D}(r)\underset{r\ll \lambda_{T}}{\simeq}-\frac{g^2c}{32\pi r^3}+\frac{g^2\pi^6rT}{\lambda^3_{T}}. \eeq
\

\paragraph*{Low temperature corrections in two dimensions.} In the same way as in the one dimensional case, there are two limiting cases for $r\ll\lambda_{T}$. For $g\ll 1$ we have:
\beq \nonumber U_{Cas}^{2D}(r,T)\underset{r\ll \lambda_{T}}{\simeq}-\frac{g^2c}{128\pi r^5}+\frac{4g^2}{3}\pi^2T^5\ln^2\big(\frac{r}{\lambda_T}\big) \eeq
At $g=1$, the leading terms are
\bea \nonumber U_{Cas}^{2D}(r,T)& &\underset{r\ll \lambda_{T}}{\simeq }-\frac{1}{2}\left(\frac{1}{r\ln\omega_cr}-\frac{1}{r\ln^2{\omega_c r}}\right)\\
\nn & &-\frac{\pi}{6}T\frac{\ln(\frac{r}{\lambda_T})}{\ln^2\omega_c^2}. \eea
\

\paragraph*{Low temperature corrections in three dimensions.} For $r\ll \lambda_{T}$, we again consider two cases. At $g=1$, we have
\beq \nonumber U_{Cas}^{3D}\underset{r\ll \lambda_{T}}{\simeq}-\frac{c}{8\pi r}Li_2\left(\frac{1}{r^2\pi^2}\right)+\frac{2T}{\pi r c}, \eeq
where $Li_2(x)$ is the polylogarithmic function.\\
And $g\ll 1$ gives us
\beq \nonumber U_{Cas}^{3D}\underset{r\ll \lambda_{T}}{\simeq}-\frac{g^2c}{256\pi^3 r^7}+\frac{2g^2\pi^2T^5}{r^2c^4}-\frac{2g^2\pi^3T^6}{rc^5}. \eeq

\paragraph*{Casimir energy at nonzero temperature in the second order of the perturbation theory.}
Here we evaluate the Casimir energy at nonzero temperature in the second order in relation to the parameter $g$. The energy is given by
\beq \nn U^{(2)}_{Cas}(r)=T\sum_{n=1}^{\infty}\left(gG^{(0)}_r(\omega_n)\right)^2, \eeq
and $G_r(\omega_n)$ is taken from Eq.(\ref{eq.GreenR.evaluated}).\\
In the one-dimensional case, the energy reads
\beq \nn U_{1D}^{(2)}=Tg^2\sum_{n=1}^{\infty}\left(\frac{\pi nT}{c}\right)^2e^{-\frac{4\pi nTr}{c}}=\frac{g^2T^3\pi^2}{4c^2}\frac{\cosh(\frac{r}{\lambda_T})}{\sinh^3(\frac{r}{\lambda_T})}.\eeq
This expression is in accordance with the result given in the Supplemental Material of \cite{Schecter2014}.\\
In the three-dimensional case, the energy turns into
\bea \nn U_{3D}^{(2)}&=&Tg^2\sum_{n=1}^{\infty}\left(\frac{\pi n^2T^2}{rc^2}\right)^2e^{-2\frac{r}{\lambda_T}}\\
\nn &=&\frac{g^2\pi^2T^5}{4r^2c^4}\frac{\cosh^3(\frac{r}{\lambda_T})+2\cosh(\frac{r}{\lambda_T})}{\sinh^5(\frac{r}{\lambda_T})}. \eea

\section{Appendix D. Infinitely heavy dynamical impurities versus impurities in an external potential}\label{appD}
In this appendix, we demonstrate the identity (\ref{ident}). For a system with dynamical impurities, it means that their masses $M\rightarrow \infty$. For a system with impurities in the external potential, it means $g\omega_0\rightarrow \infty$, so in both cases the impurities are completely static. We consider our systems on a lattice with a general spectrum and dimensionality. \\
For the $\pi\bar{\pi}$ interaction, the Green's functions are given by Eq.(\ref{GreenLatt}). For the $\varphi\bar{\varphi}$ interaction, the corresponding Green's functions are
\bea \nn \tilde{G}^{(0)}_{\mb{r}}(\omega_n)&=&V^D_c\int_{BZ}\frac{d^D\mb{k}}{(2\pi)^D}\cos\mb{kr}\frac{\omega_0^2}{\omega_n^2+\omega^2_{\mb{k}}}, \\
\nn \tilde{G}^{(0)}(\omega_n)&=&V^D_c\int_{BZ}\frac{d^D\mb{k}}{(2\pi)^D}\frac{\omega^2_0}{\omega_n^2+\omega^2_{\mb{k}}}.
 \eea
 The $T_1$- and $\tilde{T}_1$-matrices are defined by Eq.(\ref{DTmatr}) and Eq.(\ref{TTmatr}) respectively. In the limit of static impurities they take form
 \bea \nn T_1\left. (\omega_n)\right\vert_{g=1}&=&\frac{1}{1-V^D_c\int_{BZ}\frac{d^D\mb{k}}{(2\pi)^D}\left(1-\frac{\omega^2_{n}}{\omega_n^2+\omega^2_{\mb{k}}}\right)}\\
 \nn &=&\frac{1}{V^D_c\int_{BZ}\frac{d^D\mb{k}}{(2\pi)^D}\frac{\omega^2_{n}}{\omega_n^2+\omega^2_{\mb{k}}}} ,\\
\nn \tilde{T}_1\left. (\omega_n)\right\vert_{g=\infty}&=&-\frac{1}{V^D_c\int_{BZ}\frac{d^D\mb{k}}{(2\pi)^D}\frac{\omega^2_{0}}{\omega_n^2+\omega^2_{\mb{k}}}}.  \eea

 Multiplying them by $G^{(0)}_r(\omega_n)$ and $\tilde{G}^{(0)}_r(\omega_n)$ respectively, we have
 \bea \nn T_1\left.(\omega_n)\right\vert_{g=1}G^{(0)}_r(\omega_n)&=&-\frac{\int_{BZ}d^D\mb{k}\frac{\cos\mb{kr}}{\omega_n^2+\omega^2_{\mb{k}}}}{\int_{BZ}d^D\mb{k}\frac{1}{\omega_n^2+\omega^2_{\mb{k}}}} ,\\
\nn \tilde{T}_1\left.(\omega_n)\right\vert_{g=\infty}\tilde{G}^{(0)}_r(\omega_n)&=&-\frac{\int_{BZ}d^D\mb{k}\frac{\cos\mb{kr}}{\omega_n^2+\omega^2_{\mb{k}}}}{\int_{BZ}d^D\mb{k}\frac{1}{\omega_n^2+\omega^2_{\mb{k}}}}. 
 \eea  
Therefore, these two expressions are identical. 

\section{Appendix E. Proof of Equation (\ref{ident})}\label{appD}
In this appendix, we consider effects of the weak non-linearities in the spectrum of phonons. \\
For this reason, we compare the $1D$ lattice model with a continuum limit at $T=0$. The spectrum in the lattice model is $\omega_k=2c|\sin\frac{k}{2}|$. We put the lattice constant $\delta=1$.\\
Considering dynamical impurities ($\pi\bar{\pi}$-interaction), we need the Green's functions for this spectrum. These Green's functions were found in \citep{Pavlov2018}:
\bea \nn G^{(0)}_r(\omega_n)&=&-\frac{\frac{\omega_n}{2c}}{\sqrt{1+\left(\frac{\omega_n}{2c}\right)^2}}\left(\frac{\omega_n}{2c}+\sqrt{1+\left(\frac{\omega_n}{2c}\right)^2} \right)^{-2r}, \\
\nn G^{(0)}(\omega_n)&=&1-\frac{\frac{\omega_n}{2c}}{\sqrt{1+\left(\frac{\omega_n}{2c}\right)^2}}. \eea 
In the low energy limit, these expressions can be simplified as
\bea \nn G^{(0)}_r(\omega_n)&\simeq &-\frac{\omega_n}{2c}e^{-\frac{\omega_n}{c}r},\\
\nn G^{(0)}(\omega_n)&\simeq &1-\frac{\omega_n}{2c}, \eea 
giving us the Green's function for the linear spectrum.\\
For localized impurities ($\varphi\bar{\varphi}$-interaction), the corresponding Green's functions read
\bea \nn \tilde{G}^{(0)}_r(\omega_n)&=&\frac{\frac{\omega^2_0}{2\omega_nc}}{\sqrt{1+\left(\frac{\omega_n}{2c}\right)^2}}\left(\frac{\omega_n}{2c}+\sqrt{1+\left(\frac{\omega_n}{2c}\right)^2} \right)^{-2r}, \\
\nn \tilde{G}^{(0)}(\omega_n)&=&\frac{\frac{\omega^2_0}{2\omega_nc}}{\sqrt{1+\left(\frac{\omega_n}{2c}\right)^2}}. \eea 
The low energy limit Green's functions (corresponding to the case of linear spectrum) are
\bea \nn \tilde{G}^{(0)}_r(\omega_n)&\simeq &\frac{\omega^2_0}{2\omega_nc}e^{-\frac{\omega_n}{c}r}, \\
\nn \tilde{G}^{(0)}(\omega_n)&\simeq &\frac{\omega^2_0}{2\omega_nc}. \eea 
The ratios between the potentials for continuous ($U_{cont}(r)$) and lattice ($U_{lattice}(r)$) spectra are given in Fig.\ref{fig:relation}. It demonstrates that small corrections to the result ($\sim 1\%$ of potential for the linear spectrum) appear at $r\sim \delta$ and quickly vanish.

\end{document}